\documentclass[aps,prb,reprint,superscriptaddress,longbibliography,nofootinbib]{revtex4-2}

\usepackage{graphicx,amsmath,amssymb,amsfonts,dcolumn,dsfont,latexsym,color,bm,hyperref,siunitx,mathrsfs,braket, lipsum}
\usepackage[utf8]{inputenc}

\usepackage{orcidlink}

\hypersetup{
    colorlinks=true,
    linkcolor=blue,
    citecolor=blue,
    urlcolor=blue,
}

\begin{document}

\title{Double-peak Majorana bound states in altermagnet--superconductor heterostructures}

\author{Pankaj Sharma\,\orcidlink{0009-0004-2718-1881}}
\affiliation{Department of Physics, Indian Institute of Technology Roorkee, Roorkee 247667, India}

\author{Narayan Mohanta\,\orcidlink{0000-0003-3188-0445}}
\affiliation{Department of Physics, Indian Institute of Technology Roorkee, Roorkee 247667, India}

\begin{abstract}
We study Majorana bound states in a planar Josephson junction in which the middle channel is a $d$-wave altermagnetic metal deposited on a proximitized two-dimensional electron gas. 
In the topological regime, the near-zero-energy states reveals a characteristic double-peak spatial profile, with the Majorana wavefunction localized near the altermagnet–superconductor interfaces.
Using simplified theoretical models, we show that anisotropic hopping intrinsic to altermagnetism naturally generates interface-localized low-energy states, providing the natural explanation for the double-peak structure. 
In a nanowire geometry with extended normal metallic regions, the same feature persists but the Majorana bound states become more sensitive to the chemical potential compared to the case in planar Josephson junction. 
In a T-shaped Josephson junction, multiple near-zero-energy states appear, and the Majorana bound state expected at the crossing point is found to be localized near the interfaces, demonstrating that the localization of the Majorana bound states is primarily governed by interface boundaries rather than by the junction geometry. 
These results show that anisotropic hopping and interface structure play a central role in altermagnet-based topological superconductors and provide a promising route toward a network of controllable Majorana bound states without external magnetic fields.
\end{abstract}

\maketitle
 
\section{Introduction}
Topological superconductors hosting Majorana bound states (MBS) have attracted huge interest due to their non-Abelian statistics and potential application in fault-tolerant quantum computation~\cite{Kitaev_2000,Nayak_RMP2008,Alicea_2012,Lutchyn_2010,Oreg_PRL2010,Sau_PRL2010}. 
Experimental signatures consistent with MBS have been reported in semiconductor–superconductor nanowires, magnetic atom chains, and proximitized two-dimensional systems~\cite{Mourik_Science2012,Das2012,NadjPerge_Science2014,Fornieri_Nature2019,Ren_Nature2019}. 
Planar Josephson junctions formed on proximitized two-dimensional electron gases provide a versatile platform, where the topological superconducting phase can be tuned by electrostatic gating, phase bias, and device geometry~\cite{Pientka_PRX2017,Hell_PRL2017,Setiawan_PRB2019,Volpez_PRR2020, Sharma_PRB2024, Schiela_PRXQuantum2024, Banerjee_PRB2023, Banerjee_PRL2023, Sharma_PRB2025}. 
Such junctions allow the realization of multi-terminal devices and network of multiple MBS required for braiding operations. But previously-proposed realizations rely on Zeeman fields with a fixed direction or a ferromagnet that can suppress superconductivity and introduce stray magnetic fields. 
Magnetic textures such as skyrmion crystals have, therefore, been proposed as alternative ways to generate effective spin splitting without large external fields and to enable spatial control of MBS~\cite{Yang_Phys.Rev.B_2016,Rex_Phys.Rev.B_2019,Mascot_NpjQuantumMater._2021,Mohanta_CommunPhys2021, Sharma__2025}. 

\vspace{1em}

Recently, altermagnets have emerged as a new class of magnetic materials that break time-reversal symmetry while maintaining zero net magnetization~\cite{Smejkal_Phys.Rev.X_2022,Smejkal_Phys.Rev.X_2022a}. 
In contrast to conventional ferromagnets and antiferromagnets, altermagnets exhibit momentum-dependent spin splitting, leading to strong spin polarization in the band structure. 
The existence of this unconventional magnetic phase has been confirmed experimentally in several materials through angle-resolved photoemission and spectroscopic measurements~\cite{Krempasky_Nature_2024,Fedchenko_Sci.Adv._2024,Osumi_Phys.Rev.B_2024,Reimers_NatCommun_2024,Jiang_Nat.Phys._2025}. 
Because the exchange splitting in altermagnets originates from crystal symmetry rather than net magnetization, these materials provide a promising route to integrate magnetism with superconductivity without destroying the superconducting gap. 
This has motivated extensive theoretical work on superconducting and altermagnetic systems, including Josephson diode effect, unconventional pairing, spin currents, and novel transport responses~\cite{Papaj_Phys.Rev.B_2023,Amundsen_Phys.Rev.B_2024,Banerjee_Phys.Rev.B_2024,Lu_Phys.Rev.Lett._2024,Cheng_Phys.Rev.B_2024,Zhao_Phys.Rev.B_2025,Chen_Phys.Rev.B_2025,Sim_Phys.Rev.B_2025,Beenakker_Phys.Rev.B_2023,Alipourzadeh_Phys.Rev.B_2025,Fukaya_Phys.Rev.B_2025, Jabir_PRL_2023, Giil_Phys.Rev.B_2024,Monkman_Phys.Rev.X_2026}. 
The interplay of altermagnetism with superconductivity has recently been predicted to generate topological superconducting phases supporting MBS without external magnetic fields. 
In particular, altermagnetic heterostructures can host Majorana end states, chiral modes, and higher-order topological superconductivity due to the intrinsic spin splitting and anisotropic dispersion~\cite{Zhu_Phys.Rev.B_2023,Ghorashi_Phys.Rev.Lett._2024,Li_Phys.Rev.B_2023,Sun_Phys.Rev.B_2025,Yi_Phys.Rev.B_2026,Hadjipaschalis_Phys.Rev.B_2025, Mondal_Phys.Rev.B_2025,Pal_Phys.Rev.B_2025,Chatterjee_Phys.Rev.B_2025,Subhadarshini_Phys.Rev.B_2025,Yang__2025}. 
These developments establish altermagnets as a promising platform for realizing field-free topological superconductivity in planar and multi-terminal geometries.

In this work, we investigate MBS in planar Josephson junctions in which the normal channel contains a $d$-wave altermagnetic metal. 
We show that the anisotropic hopping intrinsic to altermagnetism produces low-energy states that preferentially localize near the altermagnet–normal metal interfaces, resulting in a characteristic double-peak spatial structure of the MBS in the planar Josephson junction. 
In contrast to Zeeman-field- or magnetic-texture-driven junctions, where the Majorana wavefunction typically forms a single peak at the ends of the junction, altermagnetic heterostructures naturally exhibit a double-peak profile originating from interfaces between regions with different hopping anisotropy. 
To demonstrate that this feature is not restricted to the planar geometry, we further study a quasi-nanowire with extended normal metallic regions and a T-shaped altermagnetic Josephson junction. 
Both geometries also display interface-localized double- or multi-peak Majorana profiles, showing that this behavior is robust against changes in device geometry and is a consequence of altermagnet-based topological superconducting heterostructures.


\section{Model and Method}

\begin{figure}[t]
\includegraphics[width=\linewidth]{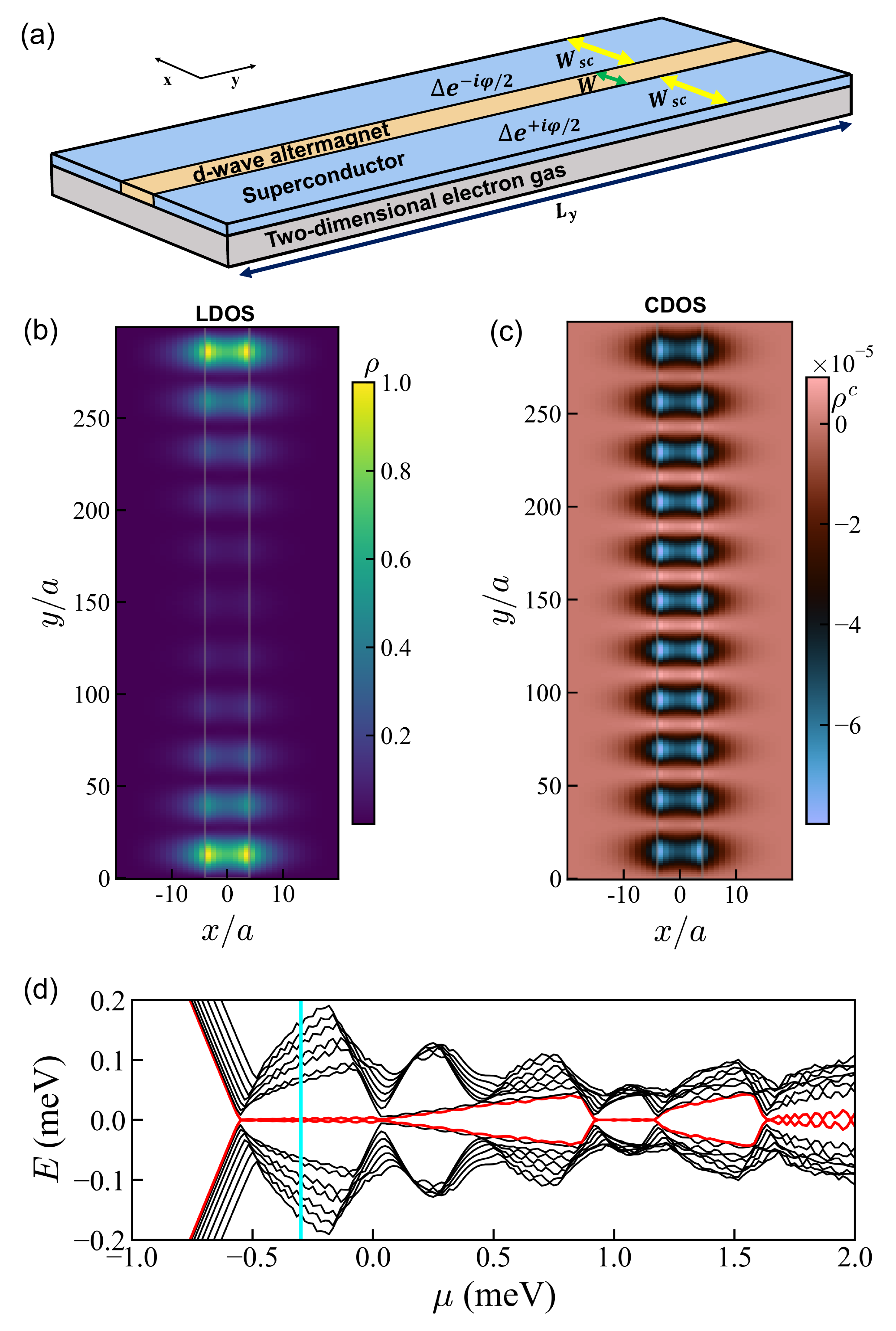}
\caption{(a) Schematic of the planar Josephson junction consisting of two $s$-wave superconducting regions separated by a $d$-wave altermagnetic channel deposited on a two-dimensional electron gas. (b) Real-space profile of the normalized local density of states and (c) charge density of states corresponding to the near-zero-energy MBS at $\mu = -0.3$~meV, indicated by the vertical cyan line in (d). (d) Quasiparticle eigenenergy spectrum of the planar Josephson junction as a function of the chemical potential.}
\label{FIG:1}
\end{figure}

We consider a two-dimensional electron gas with proximity-induced superconductivity, Rashba spin--orbit coupling, and a $d$-wave altermagnetic channel in the middle of the planar Josephson junction, as shown in Fig.~\ref{FIG:1}(a). We assume that the two-dimensional electron gas is proximitized by both, the s-wave superconductor and $d$-wave altermagnet. The superconducting pairing is induced underneath the superconducting leads, while the altermagnetic exchange field is present only in the middle channel region. The system can be described by the following tight-binding Hamiltonian
\begin{align}
\mathcal{H}
&= - t \sum_{\langle i j\rangle, \sigma}
(c^\dagger_{i\sigma} c_{j\sigma} + \mathrm{H.c.})
+ \sum_{i,\sigma} (4t-\mu) c^\dagger_{i\sigma} c_{i\sigma}
\nonumber \\
&\quad
+ \sum_i \left(
\Delta_i c^\dagger_{i\uparrow} c^\dagger_{i\downarrow}
+ \mathrm{H.c.}
\right)
- \frac{i\alpha}{2a}
\sum_{\langle i j\rangle,\sigma\sigma'}
(\boldsymbol{\sigma} \times \mathbf{d}_{ij})^z_{\sigma\sigma'}
c^\dagger_{i\sigma} c_{j\sigma'}
\nonumber \\
&\quad
- \sum_{\langle i j\rangle,\sigma\sigma'}
t_{AM} (d_x^2-d_y^2)
(\sigma_z)_{\sigma\sigma'}
c^\dagger_{i\sigma} c_{j\sigma'}
+ \mathrm{H.c.}
\label{Ham}
\end{align}
Here $t=\hbar^2/2ma^2$ is the nearest-neighbor hopping amplitude, where $m$ is the effective electron mass and $a$ is the lattice spacing of the square grid. The indices $i$ and $j$ denote lattice sites, while $\sigma,\sigma'={\{\uparrow,\downarrow}\}$ label spins. The chemical potential is denoted by $\mu$, and $\Delta_i$ is the proximity-induced $s$-wave superconducting pairing amplitude, which is finite only underneath the superconducting leads. The Rashba spin--orbit coupling strength is given by $\alpha$, $\boldsymbol{\sigma}$ represents the Pauli matrices, and $\mathbf{d}_{ij}$ is the unit vector pointing from site $i$ to site $j$.
The last term describes the altermagnetic exchange field, characterized by the hopping amplitude $t_{AM}$, which is finite only in the middle channel. The factor $(d_x^2-d_y^2)$ produces the $d$-wave anisotropic hopping associated with altermagnetism, where $d_x$ and $d_y$ denote unit vectors along the $\pm \hat{x}$ and $\pm \hat{y}$ directions, respectively.

The Hamiltonian is diagonalized using the Bogoliubov transformation
$c_{i\sigma}
=
\sum_n
\left(
u^n_{i\sigma} \gamma_n
+
v^{n*}_{i\sigma} \gamma_n^\dagger
\right),
$
which yields the quasiparticle eigenvalues and eigenvectors. Here $u^n_{i\sigma}$ and $v^n_{i\sigma}$ are the quasiparticle and quasihole amplitudes, respectively, and $\gamma_n$ ($\gamma_n^\dagger$) are fermionic annihilation (creation) operators corresponding to the $n$th Bogoliubov–de Gennes quasiparticle state.

The parameters are chosen to model experimentally relevant planar Josephson junctions formed in InAs quantum wells with Al superconducting leads. We use
$m = 0.026 m_0$ (with $m_0$ the free electron mass),
$\Delta = 0.2$~meV,
$\alpha = 30$~meV-nm,
and lattice spacing $a = 10$~nm.
The altermagnetic strength is taken as $t_{AM}=5$~meV throughout this work, unless specified otherwise.
The numerical calculations are performed using the KWANT package~\cite{Groth_NJP2014}.
The superconducting phase difference between the two leads is fixed to $\varphi = 0$.
The dimensions of the planar Josephson junction are
$L_y = 300a$, $W = 8a$, and $W_{sc} = 16a$.




\section{Results}

\textcolor{blue}{\textit{Altermagnetic planar Josephson junction:}}
We identify the topological superconducting phase in the planar Josephson junction through the emergence of near-zero-energy MBS, which are protected by a finite bulk gap. The quasiparticle eigenenergy spectrum as a function of chemical potential $\mu$ is shown in Fig.~\ref{FIG:1}(d). A topological phase transition is signaled by a closing and reopening of the bulk gap, followed by the appearance of near-zero-energy MBS in the range $-0.55 \lesssim \mu \lesssim 0.02$~meV.

The localization of MBS is confirmed by local density of states (LDOS), defined as $\rho_{i} = \sum_\sigma (|u_{i \sigma}|^2 + |v_{i \sigma}|^2)$, evaluated for the lowest positive-energy eigenstate at $\mu = -0.3$~meV. As shown in Fig.~\ref{FIG:1}(b), the LDOS reveals that the MBS are localized at the two ends of the altermagnetic channel.
A key distinguishing feature of this system is the spatial structure of the MBS. Unlike conventional Zeeman-field- or magnetic-texture-driven topological superconductors, where each Majorana appears as a single localized peak, we observe a pronounced double-peak structure at each end of the junction. The MBS density is enhanced near the altermagnet--superconductor interfaces, resulting in a characteristic dumbbell-shaped profile. This behavior reflects the intrinsic interface-localization mechanism in altermagnet-based systems.
This double-peak signature is further corroborated by the charge density of states, defined as $\rho_i^c = \sum_\sigma (|u_{i \sigma}|^2 - |v_{i \sigma}|^2)$, shown in Fig.~\ref{FIG:1}(c). The corresponding profile exhibits an oscillatory, charge-density-wave-like pattern.


\vspace{0.4 em} 

\textcolor{blue}{\textit{Localization of low-energy states at Altermagnet/normal-metal interfaces:}}
To understand the origin of the double-peak Majorana structure observed in the planar Josephson junction, we first consider a minimal model that captures the essential effect of $d$-wave altermagnetism, namely anisotropic hopping. Specifically, we introduce an anisotropic hopping term in a finite square region embedded within a larger normal-metal system, described by the following Hamiltonian

\begin{align} \mathcal{H_{\rm 1}}&=- t\sum_{\langle i j\rangle} (c^\dagger_{i}c_{j}+ {\rm H.c.}) + \sum_{i} (4t -\mu)c^\dagger_{i}c_{i}  \nonumber\\ 
&+t_{AM}\Bigg(\sum_{\langle i j\rangle_x} c^\dagger_{i}c_{j}
- \sum_{\langle i j\rangle_y} c^\dagger_{i}c_{j} + {\rm H.c.}\Bigg)
\label{Ham2}
\end{align}

\begin{figure}[t]
\includegraphics[width=\linewidth]{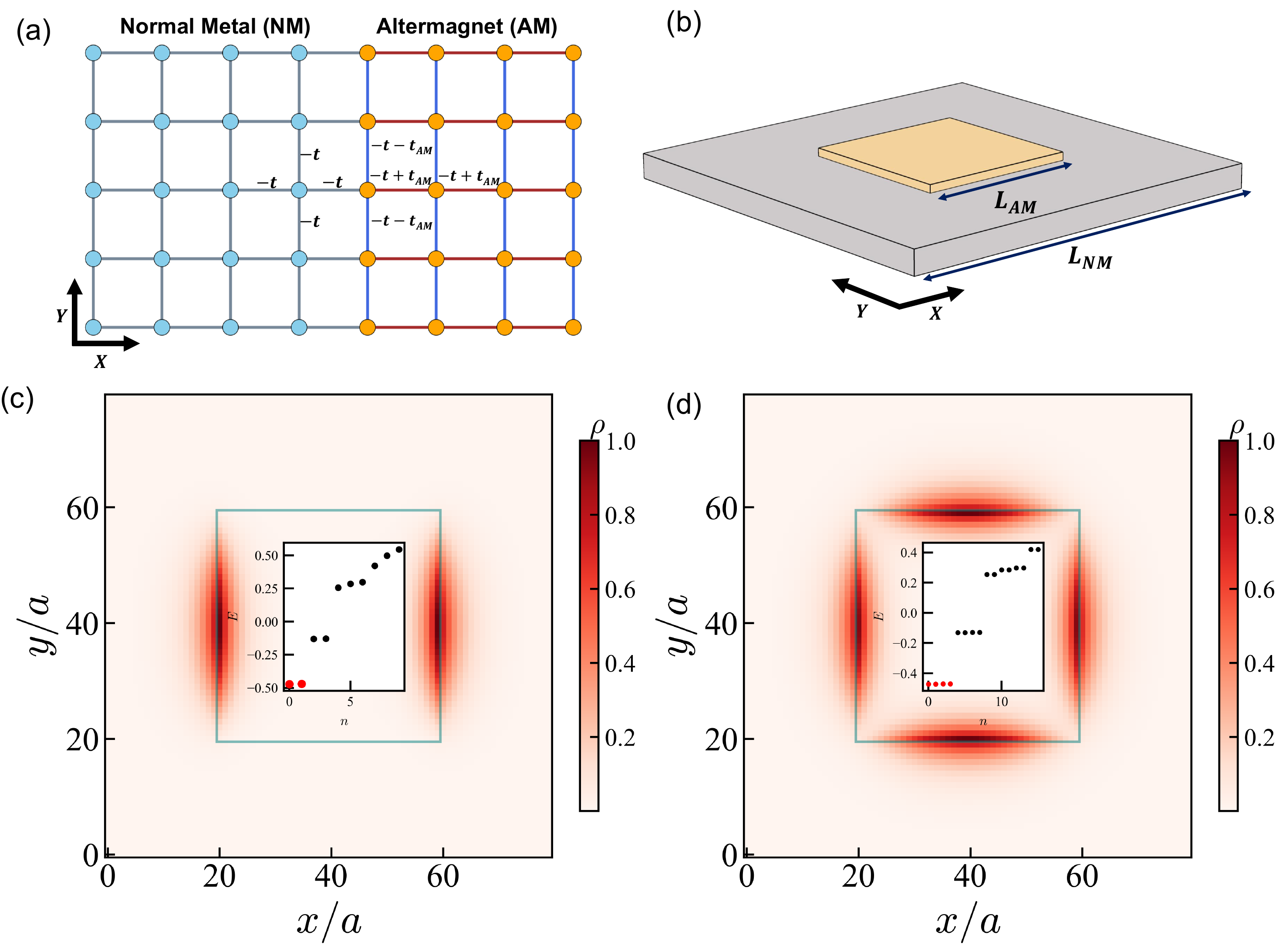}
\caption{(a) Schematic illustrating the anisotropic hopping for a single-spin channel at the altermagnet/normal-metal interface, where $t_{AM}$ denotes the altermagnetic contribution to the hopping amplitude. 
(b) Geometry of the $d$-wave altermagnet--normal-metal heterostructure, consisting of a finite altermagnetic region embedded within a larger normal-metal system. 
(c) Local density of states corresponding to the two lowest-energy degenerate eigenstates of the minimal model $\mathcal{H}_{1}$, showing localization along the interface. The inset displays the corresponding low-energy spectrum indexed by $n$. 
(d) Local density of states corresponding to the four lowest-energy degenerate eigenstates of the AM/NM heterostructure described by $\mathcal{H}_{\rm AM/NM}$, exhibiting multiple interface-localized regions. The inset shows the corresponding low-energy spectrum. 
The system dimensions are $L_{AM} = 40a$ and $L_{NM} = 80a$, with $t_{AM} = 0.3\,t$.}
\label{FIG:2}
\end{figure}

For the calculations presented here, we set $t_{AM} = 0.3t$ and $\mu = 0$. The corresponding interface between the anisotropic and isotropic regions is illustrated in Fig.~\ref{FIG:2}(a). 
The low-energy spectrum of this model, shown in the inset of Fig.~\ref{FIG:2}(c), exhibits two degenerate ground states. The associated LDOS, plotted in Fig.~\ref{FIG:2}(c), reveals that these states are strongly localized along the interface between the anisotropic region and the surrounding normal metal. This demonstrates that the presence of anisotropic hopping alone is sufficient to generate interface-localized low-energy modes.

We next consider a more realistic model of an altermagnet--normal-metal (AM/NM) heterostructure, where a finite altermagnetic region is embedded within a larger normal-metal background, as shown schematically in Fig.~\ref{FIG:2}(b). The corresponding Hamiltonian is given below 

\begin{align} \mathcal{H_{\rm AM/NM}}&=- t\sum_{\langle i j\rangle, \sigma} (c^\dagger_{i\sigma}c_{j\sigma}+ {\rm H.c.}) + \sum_{i, \sigma} (4t -\mu)c^\dagger_{i\sigma}c_{i\sigma}  \nonumber\\ 
&-\sum_{\langle ij\rangle,\sigma\sigma^\prime} t_{AM} (d_x^2-d_y^2)(\sigma_z)_{\sigma\sigma^\prime} c^\dagger_{i\sigma} c_{j\sigma^\prime} + {\rm H.c.}
\label{Ham3}
\end{align}

For the calculations presented here, we set $t_{AM} = 0.3t$ and $\mu = 0$. The low-energy spectrum of this system, shown in the inset of Fig.~\ref{FIG:2}(d), now exhibits four nearly degenerate states. The LDOS corresponding to these states, plotted in Fig.~\ref{FIG:2}(d), shows that they are localized along the AM/NM interfaces, reflecting the geometry of the interface.
These results demonstrate that low-energy states in altermagnet--normal-metal systems are intrinsically localized at interfaces where the hopping structure changes. This can be understood from the anisotropic kinetic energy induced by the altermagnetic term: the hopping amplitudes differ along the $x$ and $y$ directions, leading to direction-dependent dispersion. As a result, quasiparticles preferentially delocalize along directions with larger effective hopping while remaining confined along directions with reduced hopping. This produces states that are extended along the interfaces.

\begin{figure}[t]
\includegraphics[width=\linewidth]{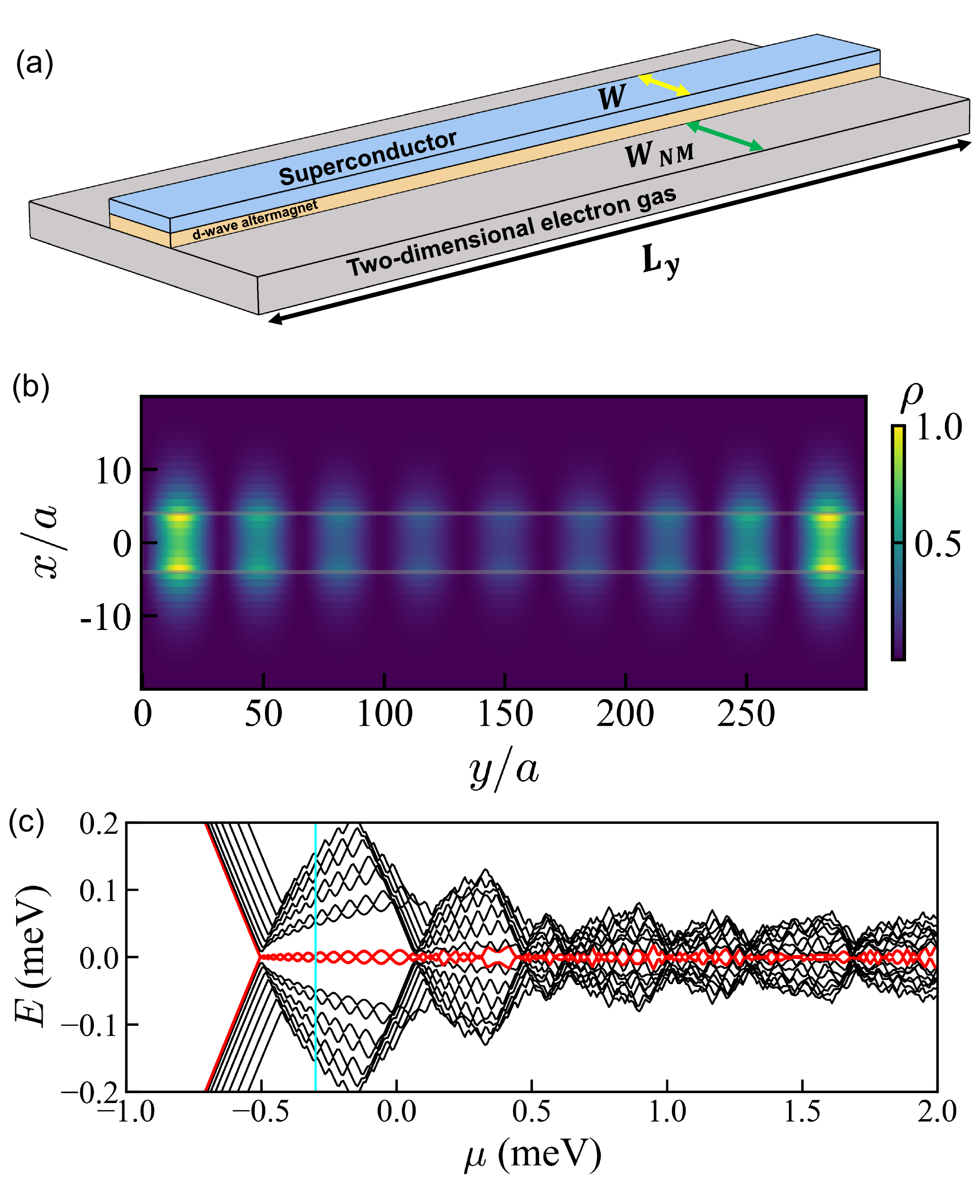}
\caption{(a) Schematic of the nanowire structure with extended normal metallic regions on both sides of the altermagnetic segment. 
(b) Real-space profile of the LDOS corresponding to the near-zero-energy MBS at $\mu = -0.3~\mathrm{meV}$ (indicated by the vertical cyan line in panel (c)). 
(c) Quasiparticle eigenenergy spectrum as a function of chemical potential for the nanowire geometry shown in (a).}
\label{FIG:3}
\end{figure}

\vspace{0.4 em} 
\textcolor{blue}{\textit{Topological nanowire with extended normal metallic regions:}}
We next investigate a quasi-one-dimensional nanowire geometry incorporating extended normal metallic regions on both sides of the altermagnetic segment, as shown in Fig.~\ref{FIG:3}(a). The dimensions of the system are chosen to match those of the planar Josephson junction discussed above, namely $W = 8a$, $W_{NM} = 16a$, and $L_y = 300a$, allowing for a direct comparison between the two geometries.

The quasiparticle eigenenergy spectrum as a function of the chemical potential is presented in Fig.~\ref{FIG:3}(c). We observe a closing and reopening of the bulk gap near $\mu \approx -0.5~\mathrm{meV}$, signaling a topological phase transition accompanied by the emergence of near-zero-energy states. In contrast to the planar Josephson junction [Fig.~\ref{FIG:1}(d)], no additional topological regions are observed over the considered parameter range.

The spatial profile of the corresponding low-energy state, shown in Fig.~\ref{FIG:3}(b), reveals a double-peak structure in the LDOS consistent with the MBS behavior identified in the planar Josephson junction geometry. However, despite having identical parameters and system dimensions, the near-zero-energy states in the nanowire exhibit stronger oscillations as a function of chemical potential [compare Fig~\ref{FIG:1}(d) and Fig~\ref{FIG:3}(c)]. This indicates that the MBS in nanowires have extended wavefunctions compared to the MBS in the planar Josephson junction, suggesting that the extended interfaces in the planar geometry provide more robust and well-localized MBS in altermagnet-based systems.

\begin{figure}[t]
\includegraphics[width=\linewidth]{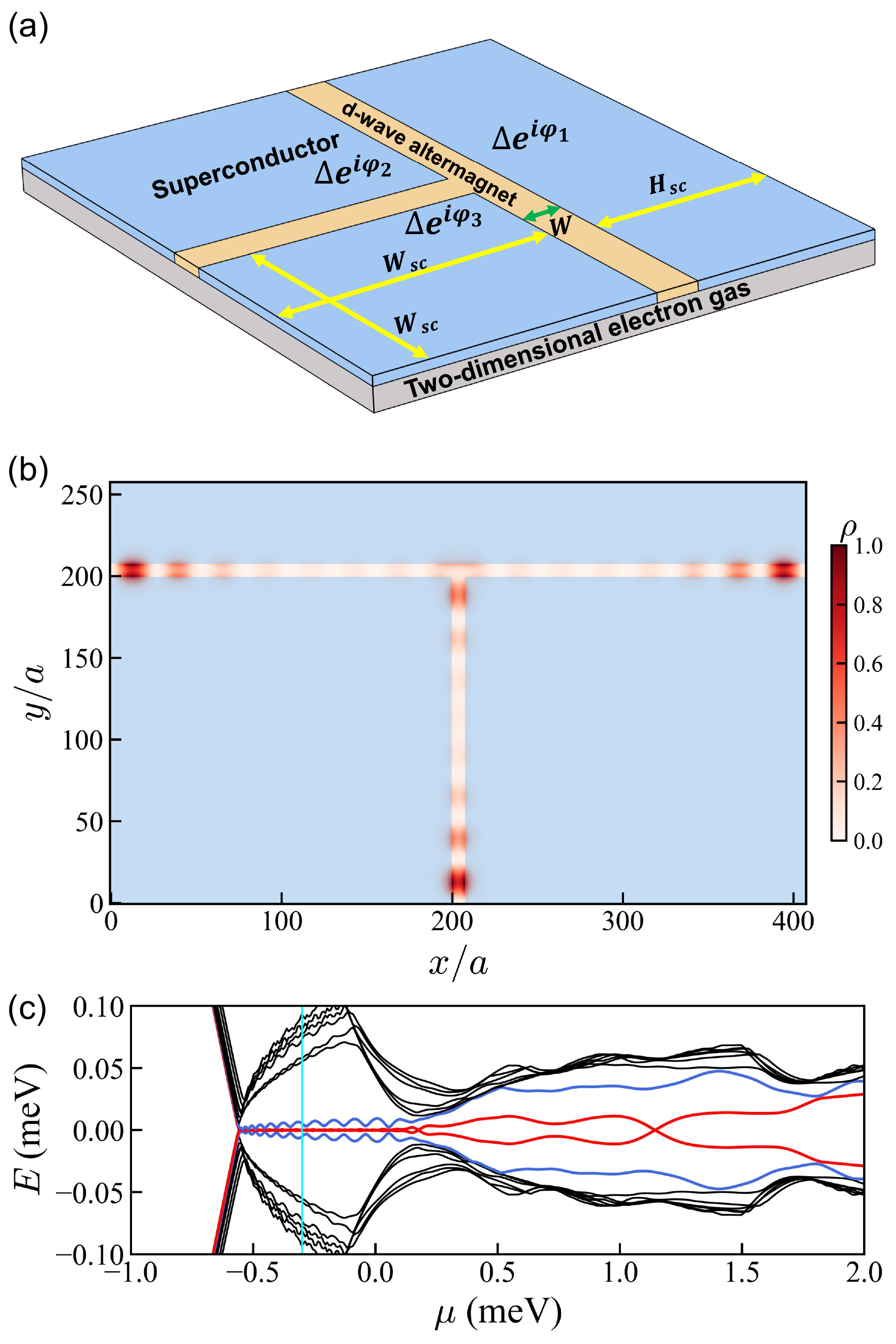}
\caption{(a) Schematic of a $T$-shaped Josephson junction in which a $d$-wave altermagnetic channel forms the vertical and horizontal arms, separating three $s$-wave superconducting regions on a two-dimensional electron gas. 
(b) Real-space profile of the LDOS corresponding to the near-zero-energy MBS at $\mu = -0.3~\mathrm{meV}$ (indicated by the vertical cyan line in panel (c)). 
(c) Quasiparticle eigenenergy spectrum as a function of chemical potential, showing the emergence of two near-zero-energy states well separated from the bulk continuum. 
The system dimensions are $W_{sc} = 200a$, $W = 8a$, and $H_{sc} = 50a$. All other parameters are identical to those used for the planar Josephson junction in Fig.~\ref{FIG:1}(a).}
\label{FIG:4}
\end{figure}

\textcolor{blue}{\textit{T-shaped altermagnetic Josephson junction:}}
Finally, we study a T-shaped Josephson junction geometry, which is relevant for multi-terminal superconducting devices and for the manipulation of MBS. The structure is shown in Fig.~\ref{FIG:4}(a), where a $d$-wave altermagnetic channel forms both the horizontal and vertical arms of the junction, separating three superconducting regions deposited on top of a two-dimensional electron gas. The superconducting phase in all three superconducting regions is taken to be identical. The widths of the altermagnetic channels are chosen to be the same as in the planar Josephson junction discussed above.
The quasiparticle eigenenergy spectrum as a function of the chemical potential is shown in Fig.~\ref{FIG:4}(c). We observe the closing and reopening of the bulk gap followed by the appearance of two near-zero-energy MBS that remain well separated from the bulk states.

From the topology of the T-junction we expect four MBS: one localized at each of the three outer ends of the junction and one around the central crossing point where the three channels meet. To verify this, we calculate the LDOS corresponding to the two lowest-energy quasiparticle eigenstates, shown in Fig.~\ref{FIG:4}(b). We indeed observe four localized regions of enhanced LDOS, but their spatial distribution differs from the initial expectation. The three states located at the outer ends of the horizontal and vertical arms exhibit the same double-peak structure as discussed earlier for the planar Josephson junction, where each Majorana mode is split into two maxima localized near the altermagnet--superconductor interfaces. This confirms that the double-peak localization is an intrinsic feature of altermagnet-based topological superconducting heterostructures and persists even in multi-terminal geometries.
The Majorana bound state associated with the central region, however, does not localize at the exact crossing point of the T-junction. Instead, the LDOS shows that this state shifts toward the nearby interfaces between the altermagnetic channel and the superconducting regions. This behavior is consistent with the interface-localization mechanism discussed in the previous sections, where low-energy states in altermagnet--normal or altermagnet--superconductor heterostructures tend to accumulate near boundaries where the effective hopping anisotropy changes. As a result, even in a geometry where a Majorana bound state is expected at the crossing point, the actual localization is controlled by the interface structure rather than by the geometric center of the junction.

\section{Discussion}

In this work, we investigated the localization of the MBS in altermagnet--superconductor heterostructures with emphasis on the role of anisotropic hopping and interface geometry. 
In contrast to conventional Zeeman-field-driven topological superconductors, where MBS typically localize at the outermost ends of the system, we find that altermagnet-based Josephson junctions naturally support low-energy states that preferentially localize near the interfaces between regions with different hopping structures. 
This behavior originates from the $d$-wave anisotropic hopping induced by the altermagnetic exchange field, which modifies the effective kinetic energy along different directions and favors localization along boundaries where the hopping anisotropy changes.

A direct consequence of this mechanism is the appearance of a double-peak Majorana structure in planar altermagnetic Josephson junctions. 
Instead of a single localized peak at each end of the junction, the Majorana wavefunction splits into two maxima located near the altermagnet--superconductor interfaces. 
By analyzing simplified altermagnet--normal-metal models, we demonstrated that anisotropic hopping alone produces interface-localized low-energy states, which naturally evolve into the double-peak MBS once superconductivity drives the system into a topological phase. 
This shows that the double-peak structure is not a fine-tuned feature of a particular geometry, but an intrinsic consequence of the altermagnetic band anisotropy and the presence of interfaces.

We further showed that the spatial profile of the MBS depends sensitively on the presence of extended normal metallic regions, which are unavoidable in realistic experimental devices. 
In nanowire geometries containing finite normal segments, near-zero-energy states still appear, but they exhibit stronger oscillations as a function of chemical potential compared to the planar Josephson junction. 
This indicates that extended interfaces in planar geometries provide more robust and well-separated MBS in altermagnet-based systems. 
These results highlight that realistic device geometry must be taken into account when designing altermagnet–superconductor heterostructures for topological superconductivity.

The interface-localization mechanism becomes even more evident in the T-shaped Josephson junction geometry studied here. 
From simple topological arguments, we expect four MBS in this structure, located at the three outer ends and at the central crossing point. 
While our calculations indeed show four localized low-energy states, the state associated with the central region does not appear at the geometric intersection. 
Instead, it shifts toward the nearby altermagnet--superconductor interfaces and spreads along the boundary. 
The other three modes retain the double-peak structure observed in the planar junction. 
These results confirm that, in altermagnet-based heterostructures, the position of the MBS is controlled primarily by interface boundaries rather than by the geometric center of the device.
The T-junction geometry is particularly relevant for Majorana manipulation and braiding protocols.
The present results show that altermagnetic Josephson junctions offer an alternative route in which the position of MBS can be tuned through electrostatic controls, without requiring strong external fields, earlier proposals use a skyrmion crystal to demonstrate the movement of Majorana bound states in T-Junction platforms~\cite{Sharma__2025}.

Overall, our results demonstrate that the double-peak Majorana bound state is an intrinsic and robust feature of altermagnet--superconductor heterostructures, arising from the interface-driven localization imposed by the anisotropic hopping of the altermagnetic phase. 
This interface localization provides a characteristic signature of MBS generated in altermagnet-based platforms.

\section*{Acknowledgments}
PS acknowledges support from the Ministry of Education, India via a research fellowship. This research is supported by Science and Engineering Research Board, India and SRIC office, IIT Roorkee (grant No. SRG/2023/001188 and IITR/SRIC/2116/FIG). 

\bibliography{Ref}

\begin{thebibliography}{58}%
\makeatletter
\providecommand \@ifxundefined [1]{%
 \@ifx{#1\undefined}
}%
\providecommand \@ifnum [1]{%
 \ifnum #1\expandafter \@firstoftwo
 \else \expandafter \@secondoftwo
 \fi
}%
\providecommand \@ifx [1]{%
 \ifx #1\expandafter \@firstoftwo
 \else \expandafter \@secondoftwo
 \fi
}%
\providecommand \natexlab [1]{#1}%
\providecommand \enquote  [1]{``#1''}%
\providecommand \bibnamefont  [1]{#1}%
\providecommand \bibfnamefont [1]{#1}%
\providecommand \citenamefont [1]{#1}%
\providecommand \href@noop [0]{\@secondoftwo}%
\providecommand \href [0]{\begingroup \@sanitize@url \@href}%
\providecommand \@href[1]{\@@startlink{#1}\@@href}%
\providecommand \@@href[1]{\endgroup#1\@@endlink}%
\providecommand \@sanitize@url [0]{\catcode `\\12\catcode `\$12\catcode
  `\&12\catcode `\#12\catcode `\^12\catcode `\_12\catcode `\%12\relax}%
\providecommand \@@startlink[1]{}%
\providecommand \@@endlink[0]{}%
\providecommand \url  [0]{\begingroup\@sanitize@url \@url }%
\providecommand \@url [1]{\endgroup\@href {#1}{\urlprefix }}%
\providecommand \urlprefix  [0]{URL }%
\providecommand \Eprint [0]{\href }%
\providecommand \doibase [0]{https://doi.org/}%
\providecommand \selectlanguage [0]{\@gobble}%
\providecommand \bibinfo  [0]{\@secondoftwo}%
\providecommand \bibfield  [0]{\@secondoftwo}%
\providecommand \translation [1]{[#1]}%
\providecommand \BibitemOpen [0]{}%
\providecommand \bibitemStop [0]{}%
\providecommand \bibitemNoStop [0]{.\EOS\space}%
\providecommand \EOS [0]{\spacefactor3000\relax}%
\providecommand \BibitemShut  [1]{\csname bibitem#1\endcsname}%
\let\auto@bib@innerbib\@empty
\bibitem [{\citenamefont {Kitaev}(2001)}]{Kitaev_2000}%
  \BibitemOpen
  \bibfield  {author} {\bibinfo {author} {\bibfnamefont {A.}~\bibnamefont
  {Kitaev}},\ }\bibfield  {title} {\bibinfo {title} {{Unpaired {M}ajorana
  fermions in quantum wires}},\ }\href
  {https://doi.org/10.1070/1063-7869/44/10S/S29} {\bibfield  {journal}
  {\bibinfo  {journal} {Phys. Usp.}\ }\textbf {\bibinfo {volume} {44}},\
  \bibinfo {pages} {131} (\bibinfo {year} {2001})}\BibitemShut {NoStop}%
\bibitem [{\citenamefont {Nayak}\ \emph {et~al.}(2008)\citenamefont {Nayak},
  \citenamefont {Simon}, \citenamefont {Stern}, \citenamefont {Freedman},\ and\
  \citenamefont {Das~Sarma}}]{Nayak_RMP2008}%
  \BibitemOpen
  \bibfield  {author} {\bibinfo {author} {\bibfnamefont {C.}~\bibnamefont
  {Nayak}}, \bibinfo {author} {\bibfnamefont {S.~H.}\ \bibnamefont {Simon}},
  \bibinfo {author} {\bibfnamefont {A.}~\bibnamefont {Stern}}, \bibinfo
  {author} {\bibfnamefont {M.}~\bibnamefont {Freedman}},\ and\ \bibinfo
  {author} {\bibfnamefont {S.}~\bibnamefont {Das~Sarma}},\ }\bibfield  {title}
  {\bibinfo {title} {Non-{A}belian anyons and topological quantum
  computation},\ }\href {https://doi.org/10.1103/RevModPhys.80.1083} {\bibfield
   {journal} {\bibinfo  {journal} {Rev. Mod. Phys.}\ }\textbf {\bibinfo
  {volume} {80}},\ \bibinfo {pages} {1083} (\bibinfo {year}
  {2008})}\BibitemShut {NoStop}%
\bibitem [{\citenamefont {Alicea}(2012)}]{Alicea_2012}%
  \BibitemOpen
  \bibfield  {author} {\bibinfo {author} {\bibfnamefont {J.}~\bibnamefont
  {Alicea}},\ }\bibfield  {title} {\bibinfo {title} {New directions in the
  pursuit of {M}ajorana fermions in solid state systems},\ }\href
  {https://doi.org/10.1088/0034-4885/75/7/076501} {\bibfield  {journal}
  {\bibinfo  {journal} {Rep. Prog. Phys.}\ }\textbf {\bibinfo {volume} {75}},\
  \bibinfo {pages} {076501} (\bibinfo {year} {2012})}\BibitemShut {NoStop}%
\bibitem [{\citenamefont {Lutchyn}\ \emph {et~al.}(2010)\citenamefont
  {Lutchyn}, \citenamefont {Sau},\ and\ \citenamefont {Das~S.}}]{Lutchyn_2010}%
  \BibitemOpen
  \bibfield  {author} {\bibinfo {author} {\bibfnamefont {R.~M.}\ \bibnamefont
  {Lutchyn}}, \bibinfo {author} {\bibfnamefont {J.~D.}\ \bibnamefont {Sau}},\
  and\ \bibinfo {author} {\bibfnamefont {S.}~\bibnamefont {Das~S.}},\
  }\bibfield  {title} {\bibinfo {title} {Majorana fermions and a topological
  phase transition in semiconductor-superconductor heterostructures},\ }\href
  {https://doi.org/10.1103/PhysRevLett.105.077001} {\bibfield  {journal}
  {\bibinfo  {journal} {Phys. Rev. Lett.}\ }\textbf {\bibinfo {volume} {105}},\
  \bibinfo {pages} {077001} (\bibinfo {year} {2010})}\BibitemShut {NoStop}%
\bibitem [{\citenamefont {Oreg}\ \emph {et~al.}(2010)\citenamefont {Oreg},
  \citenamefont {Refael},\ and\ \citenamefont {von Oppen}}]{Oreg_PRL2010}%
  \BibitemOpen
  \bibfield  {author} {\bibinfo {author} {\bibfnamefont {Y.}~\bibnamefont
  {Oreg}}, \bibinfo {author} {\bibfnamefont {G.}~\bibnamefont {Refael}},\ and\
  \bibinfo {author} {\bibfnamefont {F.}~\bibnamefont {von Oppen}},\ }\bibfield
  {title} {\bibinfo {title} {Helical liquids and {M}ajorana bound states in
  quantum wires},\ }\href {https://doi.org/10.1103/PhysRevLett.105.177002}
  {\bibfield  {journal} {\bibinfo  {journal} {Phys. Rev. Lett.}\ }\textbf
  {\bibinfo {volume} {105}},\ \bibinfo {pages} {177002} (\bibinfo {year}
  {2010})}\BibitemShut {NoStop}%
\bibitem [{\citenamefont {Sau}\ \emph {et~al.}(2010)\citenamefont {Sau},
  \citenamefont {Lutchyn}, \citenamefont {Tewari},\ and\ \citenamefont
  {Das~Sarma}}]{Sau_PRL2010}%
  \BibitemOpen
  \bibfield  {author} {\bibinfo {author} {\bibfnamefont {J.~D.}\ \bibnamefont
  {Sau}}, \bibinfo {author} {\bibfnamefont {R.~M.}\ \bibnamefont {Lutchyn}},
  \bibinfo {author} {\bibfnamefont {S.}~\bibnamefont {Tewari}},\ and\ \bibinfo
  {author} {\bibfnamefont {S.}~\bibnamefont {Das~Sarma}},\ }\bibfield  {title}
  {\bibinfo {title} {Generic new platform for topological quantum computation
  using semiconductor heterostructures},\ }\href
  {https://doi.org/10.1103/PhysRevLett.104.040502} {\bibfield  {journal}
  {\bibinfo  {journal} {Phys. Rev. Lett.}\ }\textbf {\bibinfo {volume} {104}},\
  \bibinfo {pages} {040502} (\bibinfo {year} {2010})}\BibitemShut {NoStop}%
\bibitem [{\citenamefont {Mourik}\ \emph {et~al.}(2012)\citenamefont {Mourik},
  \citenamefont {Zuo}, \citenamefont {Frolov}, \citenamefont {Plissard},
  \citenamefont {Bakkers},\ and\ \citenamefont
  {Kouwenhoven}}]{Mourik_Science2012}%
  \BibitemOpen
  \bibfield  {author} {\bibinfo {author} {\bibfnamefont {V.}~\bibnamefont
  {Mourik}}, \bibinfo {author} {\bibfnamefont {K.}~\bibnamefont {Zuo}},
  \bibinfo {author} {\bibfnamefont {S.~M.}\ \bibnamefont {Frolov}}, \bibinfo
  {author} {\bibfnamefont {S.~R.}\ \bibnamefont {Plissard}}, \bibinfo {author}
  {\bibfnamefont {E.~P. A.~M.}\ \bibnamefont {Bakkers}},\ and\ \bibinfo
  {author} {\bibfnamefont {L.~P.}\ \bibnamefont {Kouwenhoven}},\ }\bibfield
  {title} {\bibinfo {title} {Signatures of {M}ajorana fermions in hybrid
  superconductor-semiconductor nanowire devices},\ }\href
  {https://doi.org/10.1126/science.1222360} {\bibfield  {journal} {\bibinfo
  {journal} {Science}\ }\textbf {\bibinfo {volume} {336}},\ \bibinfo {pages}
  {1003} (\bibinfo {year} {2012})}\BibitemShut {NoStop}%
\bibitem [{\citenamefont {Das}\ \emph {et~al.}(2012)\citenamefont {Das},
  \citenamefont {Ronen}, \citenamefont {Most}, \citenamefont {Oreg},
  \citenamefont {Heiblum},\ and\ \citenamefont {Shtrikman}}]{Das2012}%
  \BibitemOpen
  \bibfield  {author} {\bibinfo {author} {\bibfnamefont {A.}~\bibnamefont
  {Das}}, \bibinfo {author} {\bibfnamefont {Y.}~\bibnamefont {Ronen}}, \bibinfo
  {author} {\bibfnamefont {Y.}~\bibnamefont {Most}}, \bibinfo {author}
  {\bibfnamefont {Y.}~\bibnamefont {Oreg}}, \bibinfo {author} {\bibfnamefont
  {M.}~\bibnamefont {Heiblum}},\ and\ \bibinfo {author} {\bibfnamefont
  {H.}~\bibnamefont {Shtrikman}},\ }\bibfield  {title} {\bibinfo {title}
  {Zero-bias peaks and splitting in an {Al--InAs} nanowire topological
  superconductor as a signature of {M}ajorana fermions},\ }\href
  {https://doi.org/10.1038/nphys2479} {\bibfield  {journal} {\bibinfo
  {journal} {Nat. Phys.}\ }\textbf {\bibinfo {volume} {8}},\ \bibinfo {pages}
  {887} (\bibinfo {year} {2012})}\BibitemShut {NoStop}%
\bibitem [{\citenamefont {Nadj-Perge}\ \emph {et~al.}(2014)\citenamefont
  {Nadj-Perge}, \citenamefont {Drozdov}, \citenamefont {Li}, \citenamefont
  {Chen}, \citenamefont {Jeon}, \citenamefont {Seo}, \citenamefont {MacDonald},
  \citenamefont {Bernevig},\ and\ \citenamefont
  {Yazdani}}]{NadjPerge_Science2014}%
  \BibitemOpen
  \bibfield  {author} {\bibinfo {author} {\bibfnamefont {S.}~\bibnamefont
  {Nadj-Perge}}, \bibinfo {author} {\bibfnamefont {I.~K.}\ \bibnamefont
  {Drozdov}}, \bibinfo {author} {\bibfnamefont {J.}~\bibnamefont {Li}},
  \bibinfo {author} {\bibfnamefont {H.}~\bibnamefont {Chen}}, \bibinfo {author}
  {\bibfnamefont {S.}~\bibnamefont {Jeon}}, \bibinfo {author} {\bibfnamefont
  {J.}~\bibnamefont {Seo}}, \bibinfo {author} {\bibfnamefont {A.~H.}\
  \bibnamefont {MacDonald}}, \bibinfo {author} {\bibfnamefont {B.~A.}\
  \bibnamefont {Bernevig}},\ and\ \bibinfo {author} {\bibfnamefont
  {A.}~\bibnamefont {Yazdani}},\ }\bibfield  {title} {\bibinfo {title}
  {Observation of {M}ajorana fermions in ferromagnetic atomic chains on a
  superconductor},\ }\href {https://doi.org/10.1126/science.1259327} {\bibfield
   {journal} {\bibinfo  {journal} {Science}\ }\textbf {\bibinfo {volume}
  {346}},\ \bibinfo {pages} {602} (\bibinfo {year} {2014})}\BibitemShut
  {NoStop}%
\bibitem [{\citenamefont {Fornieri}\ \emph {et~al.}(2019)\citenamefont
  {Fornieri}, \citenamefont {Whiticar}, \citenamefont {Setiawan}, \citenamefont
  {Portol{\'e}s}, \citenamefont {Drachmann}, \citenamefont {Keselman},
  \citenamefont {Gronin}, \citenamefont {Thomas}, \citenamefont {Wang},
  \citenamefont {Kallaher}, \citenamefont {Gardner}, \citenamefont {Berg},
  \citenamefont {Manfra}, \citenamefont {Stern}, \citenamefont {Marcus},\ and\
  \citenamefont {Nichele}}]{Fornieri_Nature2019}%
  \BibitemOpen
  \bibfield  {author} {\bibinfo {author} {\bibfnamefont {A.}~\bibnamefont
  {Fornieri}}, \bibinfo {author} {\bibfnamefont {A.~M.}\ \bibnamefont
  {Whiticar}}, \bibinfo {author} {\bibfnamefont {F.}~\bibnamefont {Setiawan}},
  \bibinfo {author} {\bibfnamefont {E.}~\bibnamefont {Portol{\'e}s}}, \bibinfo
  {author} {\bibfnamefont {A.~C.~C.}\ \bibnamefont {Drachmann}}, \bibinfo
  {author} {\bibfnamefont {A.}~\bibnamefont {Keselman}}, \bibinfo {author}
  {\bibfnamefont {S.}~\bibnamefont {Gronin}}, \bibinfo {author} {\bibfnamefont
  {C.}~\bibnamefont {Thomas}}, \bibinfo {author} {\bibfnamefont
  {T.}~\bibnamefont {Wang}}, \bibinfo {author} {\bibfnamefont {R.}~\bibnamefont
  {Kallaher}}, \bibinfo {author} {\bibfnamefont {G.~C.}\ \bibnamefont
  {Gardner}}, \bibinfo {author} {\bibfnamefont {E.}~\bibnamefont {Berg}},
  \bibinfo {author} {\bibfnamefont {M.~J.}\ \bibnamefont {Manfra}}, \bibinfo
  {author} {\bibfnamefont {A.}~\bibnamefont {Stern}}, \bibinfo {author}
  {\bibfnamefont {C.~M.}\ \bibnamefont {Marcus}},\ and\ \bibinfo {author}
  {\bibfnamefont {F.}~\bibnamefont {Nichele}},\ }\bibfield  {title} {\bibinfo
  {title} {Evidence of topological superconductivity in planar {J}osephson
  junctions},\ }\href {https://doi.org/10.1038/s41586-019-1068-8} {\bibfield
  {journal} {\bibinfo  {journal} {Nature}\ }\textbf {\bibinfo {volume} {569}},\
  \bibinfo {pages} {89} (\bibinfo {year} {2019})}\BibitemShut {NoStop}%
\bibitem [{\citenamefont {Ren}\ \emph {et~al.}(2019)\citenamefont {Ren},
  \citenamefont {Pientka}, \citenamefont {Hart}, \citenamefont {Pierce},
  \citenamefont {Kosowsky}, \citenamefont {Lunczer}, \citenamefont {Schlereth},
  \citenamefont {Scharf}, \citenamefont {Hankiewicz}, \citenamefont
  {Molenkamp}, \citenamefont {Halperin},\ and\ \citenamefont
  {Yacoby}}]{Ren_Nature2019}%
  \BibitemOpen
  \bibfield  {author} {\bibinfo {author} {\bibfnamefont {H.}~\bibnamefont
  {Ren}}, \bibinfo {author} {\bibfnamefont {F.}~\bibnamefont {Pientka}},
  \bibinfo {author} {\bibfnamefont {S.}~\bibnamefont {Hart}}, \bibinfo {author}
  {\bibfnamefont {A.~T.}\ \bibnamefont {Pierce}}, \bibinfo {author}
  {\bibfnamefont {M.}~\bibnamefont {Kosowsky}}, \bibinfo {author}
  {\bibfnamefont {L.}~\bibnamefont {Lunczer}}, \bibinfo {author} {\bibfnamefont
  {R.}~\bibnamefont {Schlereth}}, \bibinfo {author} {\bibfnamefont
  {B.}~\bibnamefont {Scharf}}, \bibinfo {author} {\bibfnamefont {E.~M.}\
  \bibnamefont {Hankiewicz}}, \bibinfo {author} {\bibfnamefont {L.~W.}\
  \bibnamefont {Molenkamp}}, \bibinfo {author} {\bibfnamefont {B.~I.}\
  \bibnamefont {Halperin}},\ and\ \bibinfo {author} {\bibfnamefont
  {A.}~\bibnamefont {Yacoby}},\ }\bibfield  {title} {\bibinfo {title}
  {Topological superconductivity in a phase-controlled {J}osephson junction},\
  }\href {https://doi.org/10.1038/s41586-019-1148-9} {\bibfield  {journal}
  {\bibinfo  {journal} {Nature}\ }\textbf {\bibinfo {volume} {569}},\ \bibinfo
  {pages} {93} (\bibinfo {year} {2019})}\BibitemShut {NoStop}%
\bibitem [{\citenamefont {Pientka}\ \emph {et~al.}(2017)\citenamefont
  {Pientka}, \citenamefont {Keselman}, \citenamefont {Berg}, \citenamefont
  {Yacoby}, \citenamefont {Stern},\ and\ \citenamefont
  {Halperin}}]{Pientka_PRX2017}%
  \BibitemOpen
  \bibfield  {author} {\bibinfo {author} {\bibfnamefont {F.}~\bibnamefont
  {Pientka}}, \bibinfo {author} {\bibfnamefont {A.}~\bibnamefont {Keselman}},
  \bibinfo {author} {\bibfnamefont {E.}~\bibnamefont {Berg}}, \bibinfo {author}
  {\bibfnamefont {A.}~\bibnamefont {Yacoby}}, \bibinfo {author} {\bibfnamefont
  {A.}~\bibnamefont {Stern}},\ and\ \bibinfo {author} {\bibfnamefont {B.~I.}\
  \bibnamefont {Halperin}},\ }\bibfield  {title} {\bibinfo {title} {Topological
  superconductivity in a planar {J}osephson junction},\ }\href
  {https://doi.org/10.1103/PhysRevX.7.021032} {\bibfield  {journal} {\bibinfo
  {journal} {Phys. Rev. X}\ }\textbf {\bibinfo {volume} {7}},\ \bibinfo {pages}
  {021032} (\bibinfo {year} {2017})}\BibitemShut {NoStop}%
\bibitem [{\citenamefont {Hell}\ \emph {et~al.}(2017)\citenamefont {Hell},
  \citenamefont {Leijnse},\ and\ \citenamefont {Flensberg}}]{Hell_PRL2017}%
  \BibitemOpen
  \bibfield  {author} {\bibinfo {author} {\bibfnamefont {M.}~\bibnamefont
  {Hell}}, \bibinfo {author} {\bibfnamefont {M.}~\bibnamefont {Leijnse}},\ and\
  \bibinfo {author} {\bibfnamefont {K.}~\bibnamefont {Flensberg}},\ }\bibfield
  {title} {\bibinfo {title} {Two-dimensional platform for networks of
  {M}ajorana bound states},\ }\href
  {https://doi.org/10.1103/PhysRevLett.118.107701} {\bibfield  {journal}
  {\bibinfo  {journal} {Phys. Rev. Lett.}\ }\textbf {\bibinfo {volume} {118}},\
  \bibinfo {pages} {107701} (\bibinfo {year} {2017})}\BibitemShut {NoStop}%
\bibitem [{\citenamefont {Setiawan}\ \emph {et~al.}(2019)\citenamefont
  {Setiawan}, \citenamefont {Stern},\ and\ \citenamefont
  {Berg}}]{Setiawan_PRB2019}%
  \BibitemOpen
  \bibfield  {author} {\bibinfo {author} {\bibfnamefont {F.}~\bibnamefont
  {Setiawan}}, \bibinfo {author} {\bibfnamefont {A.}~\bibnamefont {Stern}},\
  and\ \bibinfo {author} {\bibfnamefont {E.}~\bibnamefont {Berg}},\ }\bibfield
  {title} {\bibinfo {title} {Topological superconductivity in planar
  {J}osephson junctions: {N}arrowing down to the nanowire limit},\ }\href
  {https://doi.org/10.1103/PhysRevB.99.220506} {\bibfield  {journal} {\bibinfo
  {journal} {Phys. Rev. B}\ }\textbf {\bibinfo {volume} {99}},\ \bibinfo
  {pages} {220506} (\bibinfo {year} {2019})}\BibitemShut {NoStop}%
\bibitem [{\citenamefont {Volpez}\ \emph {et~al.}(2020)\citenamefont {Volpez},
  \citenamefont {Loss},\ and\ \citenamefont {Klinovaja}}]{Volpez_PRR2020}%
  \BibitemOpen
  \bibfield  {author} {\bibinfo {author} {\bibfnamefont {Y.}~\bibnamefont
  {Volpez}}, \bibinfo {author} {\bibfnamefont {D.}~\bibnamefont {Loss}},\ and\
  \bibinfo {author} {\bibfnamefont {J.}~\bibnamefont {Klinovaja}},\ }\bibfield
  {title} {\bibinfo {title} {Time-reversal invariant topological
  superconductivity in planar {J}osephson bijunction},\ }\href
  {https://doi.org/10.1103/PhysRevResearch.2.023415} {\bibfield  {journal}
  {\bibinfo  {journal} {Phys. Rev. Res.}\ }\textbf {\bibinfo {volume} {2}},\
  \bibinfo {pages} {023415} (\bibinfo {year} {2020})}\BibitemShut {NoStop}%
\bibitem [{\citenamefont {Sharma}\ and\ \citenamefont
  {Mohanta}(2024)}]{Sharma_PRB2024}%
  \BibitemOpen
  \bibfield  {author} {\bibinfo {author} {\bibfnamefont {P.}~\bibnamefont
  {Sharma}}\ and\ \bibinfo {author} {\bibfnamefont {N.}~\bibnamefont
  {Mohanta}},\ }\bibfield  {title} {\bibinfo {title} {Challenges in detecting
  topological superconducting transitions via supercurrent and phase probes in
  planar {J}osephson junctions},\ }\href
  {https://doi.org/10.1103/PhysRevB.109.054515} {\bibfield  {journal} {\bibinfo
   {journal} {Phys. Rev. B}\ }\textbf {\bibinfo {volume} {109}},\ \bibinfo
  {pages} {054515} (\bibinfo {year} {2024})}\BibitemShut {NoStop}%
\bibitem [{\citenamefont {Schiela}\ \emph {et~al.}(2024)\citenamefont
  {Schiela}, \citenamefont {Yu},\ and\ \citenamefont
  {Shabani}}]{Schiela_PRXQuantum2024}%
  \BibitemOpen
  \bibfield  {author} {\bibinfo {author} {\bibfnamefont {W.~F.}\ \bibnamefont
  {Schiela}}, \bibinfo {author} {\bibfnamefont {P.}~\bibnamefont {Yu}},\ and\
  \bibinfo {author} {\bibfnamefont {J.}~\bibnamefont {Shabani}},\ }\bibfield
  {title} {\bibinfo {title} {Progress in superconductor-semiconductor
  topological {J}osephson junctions},\ }\href
  {https://doi.org/10.1103/PRXQuantum.5.030102} {\bibfield  {journal} {\bibinfo
   {journal} {PRX Quantum}\ }\textbf {\bibinfo {volume} {5}},\ \bibinfo {pages}
  {030102} (\bibinfo {year} {2024})}\BibitemShut {NoStop}%
\bibitem [{\citenamefont {Banerjee}\ \emph
  {et~al.}(2023{\natexlab{a}})\citenamefont {Banerjee}, \citenamefont {Lesser},
  \citenamefont {Rahman}, \citenamefont {Wang}, \citenamefont {Li},
  \citenamefont {Kringh\o{}j}, \citenamefont {Whiticar}, \citenamefont
  {Drachmann}, \citenamefont {Thomas}, \citenamefont {Wang}, \citenamefont
  {Manfra}, \citenamefont {Berg}, \citenamefont {Oreg}, \citenamefont {Stern},\
  and\ \citenamefont {Marcus}}]{Banerjee_PRB2023}%
  \BibitemOpen
  \bibfield  {author} {\bibinfo {author} {\bibfnamefont {A.}~\bibnamefont
  {Banerjee}}, \bibinfo {author} {\bibfnamefont {O.}~\bibnamefont {Lesser}},
  \bibinfo {author} {\bibfnamefont {M.~A.}\ \bibnamefont {Rahman}}, \bibinfo
  {author} {\bibfnamefont {H.-R.}\ \bibnamefont {Wang}}, \bibinfo {author}
  {\bibfnamefont {M.-R.}\ \bibnamefont {Li}}, \bibinfo {author} {\bibfnamefont
  {A.}~\bibnamefont {Kringh\o{}j}}, \bibinfo {author} {\bibfnamefont {A.~M.}\
  \bibnamefont {Whiticar}}, \bibinfo {author} {\bibfnamefont {A.~C.~C.}\
  \bibnamefont {Drachmann}}, \bibinfo {author} {\bibfnamefont {C.}~\bibnamefont
  {Thomas}}, \bibinfo {author} {\bibfnamefont {T.}~\bibnamefont {Wang}},
  \bibinfo {author} {\bibfnamefont {M.~J.}\ \bibnamefont {Manfra}}, \bibinfo
  {author} {\bibfnamefont {E.}~\bibnamefont {Berg}}, \bibinfo {author}
  {\bibfnamefont {Y.}~\bibnamefont {Oreg}}, \bibinfo {author} {\bibfnamefont
  {A.}~\bibnamefont {Stern}},\ and\ \bibinfo {author} {\bibfnamefont {C.~M.}\
  \bibnamefont {Marcus}},\ }\bibfield  {title} {\bibinfo {title} {{Signatures
  of a topological phase transition in a planar Josephson junction}},\ }\href
  {https://doi.org/10.1103/PhysRevB.107.245304} {\bibfield  {journal} {\bibinfo
   {journal} {Phys. Rev. B}\ }\textbf {\bibinfo {volume} {107}},\ \bibinfo
  {pages} {245304} (\bibinfo {year} {2023}{\natexlab{a}})}\BibitemShut
  {NoStop}%
\bibitem [{\citenamefont {Banerjee}\ \emph
  {et~al.}(2023{\natexlab{b}})\citenamefont {Banerjee}, \citenamefont {Lesser},
  \citenamefont {Rahman}, \citenamefont {Thomas}, \citenamefont {Wang},
  \citenamefont {Manfra}, \citenamefont {Berg}, \citenamefont {Oreg},
  \citenamefont {Stern},\ and\ \citenamefont {Marcus}}]{Banerjee_PRL2023}%
  \BibitemOpen
  \bibfield  {author} {\bibinfo {author} {\bibfnamefont {A.}~\bibnamefont
  {Banerjee}}, \bibinfo {author} {\bibfnamefont {O.}~\bibnamefont {Lesser}},
  \bibinfo {author} {\bibfnamefont {M.~A.}\ \bibnamefont {Rahman}}, \bibinfo
  {author} {\bibfnamefont {C.}~\bibnamefont {Thomas}}, \bibinfo {author}
  {\bibfnamefont {T.}~\bibnamefont {Wang}}, \bibinfo {author} {\bibfnamefont
  {M.~J.}\ \bibnamefont {Manfra}}, \bibinfo {author} {\bibfnamefont
  {E.}~\bibnamefont {Berg}}, \bibinfo {author} {\bibfnamefont {Y.}~\bibnamefont
  {Oreg}}, \bibinfo {author} {\bibfnamefont {A.}~\bibnamefont {Stern}},\ and\
  \bibinfo {author} {\bibfnamefont {C.~M.}\ \bibnamefont {Marcus}},\ }\bibfield
   {title} {\bibinfo {title} {{Local and nonlocal transport spectroscopy in
  planar Josephson junctions}},\ }\href
  {https://doi.org/10.1103/PhysRevLett.130.096202} {\bibfield  {journal}
  {\bibinfo  {journal} {Phys. Rev. Lett.}\ }\textbf {\bibinfo {volume} {130}},\
  \bibinfo {pages} {096202} (\bibinfo {year} {2023}{\natexlab{b}})}\BibitemShut
  {NoStop}%
\bibitem [{\citenamefont {Sharma}\ and\ \citenamefont
  {Mohanta}(2025{\natexlab{a}})}]{Sharma_PRB2025}%
  \BibitemOpen
  \bibfield  {author} {\bibinfo {author} {\bibfnamefont {P.}~\bibnamefont
  {Sharma}}\ and\ \bibinfo {author} {\bibfnamefont {N.}~\bibnamefont
  {Mohanta}},\ }\bibfield  {title} {\bibinfo {title} {Multiple majorana bound
  states and their resilience against disorder in planar {J}osephson
  junctions},\ }\href {https://doi.org/10.1103/tl5f-18zx} {\bibfield  {journal}
  {\bibinfo  {journal} {Phys. Rev. B}\ }\textbf {\bibinfo {volume} {112}},\
  \bibinfo {pages} {134501} (\bibinfo {year} {2025}{\natexlab{a}})}\BibitemShut
  {NoStop}%
\bibitem [{\citenamefont {Yang}\ \emph {et~al.}(2016)\citenamefont {Yang},
  \citenamefont {Stano}, \citenamefont {Klinovaja},\ and\ \citenamefont
  {Loss}}]{Yang_Phys.Rev.B_2016}%
  \BibitemOpen
  \bibfield  {author} {\bibinfo {author} {\bibfnamefont {G.}~\bibnamefont
  {Yang}}, \bibinfo {author} {\bibfnamefont {P.}~\bibnamefont {Stano}},
  \bibinfo {author} {\bibfnamefont {J.}~\bibnamefont {Klinovaja}},\ and\
  \bibinfo {author} {\bibfnamefont {D.}~\bibnamefont {Loss}},\ }\bibfield
  {title} {\bibinfo {title} {Majorana bound states in magnetic skyrmions},\
  }\href {https://doi.org/10.1103/PhysRevB.93.224505} {\bibfield  {journal}
  {\bibinfo  {journal} {Phys. Rev. B}\ }\textbf {\bibinfo {volume} {93}},\
  \bibinfo {pages} {224505} (\bibinfo {year} {2016})}\BibitemShut {NoStop}%
\bibitem [{\citenamefont {Rex}\ \emph {et~al.}(2019)\citenamefont {Rex},
  \citenamefont {Gornyi},\ and\ \citenamefont {Mirlin}}]{Rex_Phys.Rev.B_2019}%
  \BibitemOpen
  \bibfield  {author} {\bibinfo {author} {\bibfnamefont {S.}~\bibnamefont
  {Rex}}, \bibinfo {author} {\bibfnamefont {I.~V.}\ \bibnamefont {Gornyi}},\
  and\ \bibinfo {author} {\bibfnamefont {A.~D.}\ \bibnamefont {Mirlin}},\
  }\bibfield  {title} {\bibinfo {title} {Majorana bound states in magnetic
  skyrmions imposed onto a superconductor},\ }\href
  {https://doi.org/10.1103/PhysRevB.100.064504} {\bibfield  {journal} {\bibinfo
   {journal} {Phys. Rev. B}\ }\textbf {\bibinfo {volume} {100}},\ \bibinfo
  {pages} {064504} (\bibinfo {year} {2019})}\BibitemShut {NoStop}%
\bibitem [{\citenamefont {Mascot}\ \emph {et~al.}(2021)\citenamefont {Mascot},
  \citenamefont {Bedow}, \citenamefont {Graham}, \citenamefont {Rachel},\ and\
  \citenamefont {Morr}}]{Mascot_NpjQuantumMater._2021}%
  \BibitemOpen
  \bibfield  {author} {\bibinfo {author} {\bibfnamefont {E.}~\bibnamefont
  {Mascot}}, \bibinfo {author} {\bibfnamefont {J.}~\bibnamefont {Bedow}},
  \bibinfo {author} {\bibfnamefont {M.}~\bibnamefont {Graham}}, \bibinfo
  {author} {\bibfnamefont {S.}~\bibnamefont {Rachel}},\ and\ \bibinfo {author}
  {\bibfnamefont {D.~K.}\ \bibnamefont {Morr}},\ }\bibfield  {title} {\bibinfo
  {title} {Topological superconductivity in skyrmion lattices},\ }\href
  {https://doi.org/10.1038/s41535-020-00299-x} {\bibfield  {journal} {\bibinfo
  {journal} {npj Quantum Mater.}\ }\textbf {\bibinfo {volume} {6}},\ \bibinfo
  {pages} {6} (\bibinfo {year} {2021})}\BibitemShut {NoStop}%
\bibitem [{\citenamefont {Mohanta}\ \emph {et~al.}(2021)\citenamefont
  {Mohanta}, \citenamefont {Okamoto},\ and\ \citenamefont
  {Dagotto}}]{Mohanta_CommunPhys2021}%
  \BibitemOpen
  \bibfield  {author} {\bibinfo {author} {\bibfnamefont {N.}~\bibnamefont
  {Mohanta}}, \bibinfo {author} {\bibfnamefont {S.}~\bibnamefont {Okamoto}},\
  and\ \bibinfo {author} {\bibfnamefont {E.}~\bibnamefont {Dagotto}},\
  }\bibfield  {title} {\bibinfo {title} {Skyrmion control of {M}ajorana states
  in planar {J}osephson junctions},\ }\href
  {https://doi.org/10.1038%2Fs42005-021-00666-5} {\bibfield  {journal}
  {\bibinfo  {journal} {Commun. Phys.}\ }\textbf {\bibinfo {volume} {4}},\
  \bibinfo {pages} {163} (\bibinfo {year} {2021})}\BibitemShut {NoStop}%
\bibitem [{\citenamefont {Sharma}\ and\ \citenamefont
  {Mohanta}(2025{\natexlab{b}})}]{Sharma__2025}%
  \BibitemOpen
  \bibfield  {author} {\bibinfo {author} {\bibfnamefont {P.}~\bibnamefont
  {Sharma}}\ and\ \bibinfo {author} {\bibfnamefont {N.}~\bibnamefont
  {Mohanta}},\ }\href@noop {} {\bibinfo {title} {Magnetic field-free braiding
  and nontrivial fusion of majorana bound states in high-temperature planar
  josephson junctions}} (\bibinfo {year} {2025}{\natexlab{b}}),\ \Eprint
  {https://arxiv.org/abs/2506.04338} {arXiv:2506.04338 [cond-mat.supr-con]}
  \BibitemShut {NoStop}%
\bibitem [{\citenamefont {{\v S}mejkal}\ \emph
  {et~al.}(2022{\natexlab{a}})\citenamefont {{\v S}mejkal}, \citenamefont
  {Sinova},\ and\ \citenamefont {Jungwirth}}]{Smejkal_Phys.Rev.X_2022}%
  \BibitemOpen
  \bibfield  {author} {\bibinfo {author} {\bibfnamefont {L.}~\bibnamefont {{\v
  S}mejkal}}, \bibinfo {author} {\bibfnamefont {J.}~\bibnamefont {Sinova}},\
  and\ \bibinfo {author} {\bibfnamefont {T.}~\bibnamefont {Jungwirth}},\
  }\bibfield  {title} {\bibinfo {title} {Beyond {{Conventional Ferromagnetism}}
  and {{Antiferromagnetism}}: {{A Phase}} with {{Nonrelativistic Spin}} and
  {{Crystal Rotation Symmetry}}},\ }\href
  {https://doi.org/10.1103/PhysRevX.12.031042} {\bibfield  {journal} {\bibinfo
  {journal} {Phys. Rev. X}\ }\textbf {\bibinfo {volume} {12}},\ \bibinfo
  {pages} {031042} (\bibinfo {year} {2022}{\natexlab{a}})}\BibitemShut
  {NoStop}%
\bibitem [{\citenamefont {{\v S}mejkal}\ \emph
  {et~al.}(2022{\natexlab{b}})\citenamefont {{\v S}mejkal}, \citenamefont
  {Sinova},\ and\ \citenamefont {Jungwirth}}]{Smejkal_Phys.Rev.X_2022a}%
  \BibitemOpen
  \bibfield  {author} {\bibinfo {author} {\bibfnamefont {L.}~\bibnamefont {{\v
  S}mejkal}}, \bibinfo {author} {\bibfnamefont {J.}~\bibnamefont {Sinova}},\
  and\ \bibinfo {author} {\bibfnamefont {T.}~\bibnamefont {Jungwirth}},\
  }\bibfield  {title} {\bibinfo {title} {Emerging {{Research Landscape}} of
  {{Altermagnetism}}},\ }\href {https://doi.org/10.1103/PhysRevX.12.040501}
  {\bibfield  {journal} {\bibinfo  {journal} {Phys. Rev. X}\ }\textbf {\bibinfo
  {volume} {12}},\ \bibinfo {pages} {040501} (\bibinfo {year}
  {2022}{\natexlab{b}})}\BibitemShut {NoStop}%
\bibitem [{\citenamefont {Krempask{\'y}}\ \emph {et~al.}(2024)\citenamefont
  {Krempask{\'y}}, \citenamefont {{\v S}mejkal}, \citenamefont {D'Souza},
  \citenamefont {Hajlaoui}, \citenamefont {Springholz}, \citenamefont
  {Uhl{\'i}{\v r}ov{\'a}}, \citenamefont {Alarab}, \citenamefont
  {Constantinou}, \citenamefont {Strocov}, \citenamefont {Usanov},
  \citenamefont {Pudelko}, \citenamefont {{Gonz{\'a}lez-Hern{\'a}ndez}},
  \citenamefont {Birk~Hellenes}, \citenamefont {Jansa}, \citenamefont
  {Reichlov{\'a}}, \citenamefont {{\v S}ob{\'a}{\v n}}, \citenamefont
  {Gonzalez~Betancourt}, \citenamefont {Wadley}, \citenamefont {Sinova},
  \citenamefont {Kriegner}, \citenamefont {Min{\'a}r}, \citenamefont {Dil},\
  and\ \citenamefont {Jungwirth}}]{Krempasky_Nature_2024}%
  \BibitemOpen
  \bibfield  {author} {\bibinfo {author} {\bibfnamefont {J.}~\bibnamefont
  {Krempask{\'y}}}, \bibinfo {author} {\bibfnamefont {L.}~\bibnamefont {{\v
  S}mejkal}}, \bibinfo {author} {\bibfnamefont {S.~W.}\ \bibnamefont
  {D'Souza}}, \bibinfo {author} {\bibfnamefont {M.}~\bibnamefont {Hajlaoui}},
  \bibinfo {author} {\bibfnamefont {G.}~\bibnamefont {Springholz}}, \bibinfo
  {author} {\bibfnamefont {K.}~\bibnamefont {Uhl{\'i}{\v r}ov{\'a}}}, \bibinfo
  {author} {\bibfnamefont {F.}~\bibnamefont {Alarab}}, \bibinfo {author}
  {\bibfnamefont {P.~C.}\ \bibnamefont {Constantinou}}, \bibinfo {author}
  {\bibfnamefont {V.}~\bibnamefont {Strocov}}, \bibinfo {author} {\bibfnamefont
  {D.}~\bibnamefont {Usanov}}, \bibinfo {author} {\bibfnamefont {W.~R.}\
  \bibnamefont {Pudelko}}, \bibinfo {author} {\bibfnamefont {R.}~\bibnamefont
  {{Gonz{\'a}lez-Hern{\'a}ndez}}}, \bibinfo {author} {\bibfnamefont
  {A.}~\bibnamefont {Birk~Hellenes}}, \bibinfo {author} {\bibfnamefont
  {Z.}~\bibnamefont {Jansa}}, \bibinfo {author} {\bibfnamefont
  {H.}~\bibnamefont {Reichlov{\'a}}}, \bibinfo {author} {\bibfnamefont
  {Z.}~\bibnamefont {{\v S}ob{\'a}{\v n}}}, \bibinfo {author} {\bibfnamefont
  {R.~D.}\ \bibnamefont {Gonzalez~Betancourt}}, \bibinfo {author}
  {\bibfnamefont {P.}~\bibnamefont {Wadley}}, \bibinfo {author} {\bibfnamefont
  {J.}~\bibnamefont {Sinova}}, \bibinfo {author} {\bibfnamefont
  {D.}~\bibnamefont {Kriegner}}, \bibinfo {author} {\bibfnamefont
  {J.}~\bibnamefont {Min{\'a}r}}, \bibinfo {author} {\bibfnamefont {J.~H.}\
  \bibnamefont {Dil}},\ and\ \bibinfo {author} {\bibfnamefont {T.}~\bibnamefont
  {Jungwirth}},\ }\bibfield  {title} {\bibinfo {title} {Altermagnetic lifting
  of {{Kramers}} spin degeneracy},\ }\href
  {https://doi.org/10.1038/s41586-023-06907-7} {\bibfield  {journal} {\bibinfo
  {journal} {Nature}\ }\textbf {\bibinfo {volume} {626}},\ \bibinfo {pages}
  {517} (\bibinfo {year} {2024})}\BibitemShut {NoStop}%
\bibitem [{\citenamefont {Fedchenko}\ \emph {et~al.}(2024)\citenamefont
  {Fedchenko}, \citenamefont {Min{\'a}r}, \citenamefont {Akashdeep},
  \citenamefont {D'Souza}, \citenamefont {Vasilyev}, \citenamefont {Tkach},
  \citenamefont {Odenbreit}, \citenamefont {Nguyen}, \citenamefont
  {Kutnyakhov}, \citenamefont {Wind}, \citenamefont {Wenthaus}, \citenamefont
  {Scholz}, \citenamefont {Rossnagel}, \citenamefont {Hoesch}, \citenamefont
  {Aeschlimann}, \citenamefont {Stadtm{\"u}ller}, \citenamefont {Kl{\"a}ui},
  \citenamefont {Sch{\"o}nhense}, \citenamefont {Jungwirth}, \citenamefont
  {Hellenes}, \citenamefont {Jakob}, \citenamefont {{\v S}mejkal},
  \citenamefont {Sinova},\ and\ \citenamefont
  {Elmers}}]{Fedchenko_Sci.Adv._2024}%
  \BibitemOpen
  \bibfield  {author} {\bibinfo {author} {\bibfnamefont {O.}~\bibnamefont
  {Fedchenko}}, \bibinfo {author} {\bibfnamefont {J.}~\bibnamefont
  {Min{\'a}r}}, \bibinfo {author} {\bibfnamefont {A.}~\bibnamefont
  {Akashdeep}}, \bibinfo {author} {\bibfnamefont {S.~W.}\ \bibnamefont
  {D'Souza}}, \bibinfo {author} {\bibfnamefont {D.}~\bibnamefont {Vasilyev}},
  \bibinfo {author} {\bibfnamefont {O.}~\bibnamefont {Tkach}}, \bibinfo
  {author} {\bibfnamefont {L.}~\bibnamefont {Odenbreit}}, \bibinfo {author}
  {\bibfnamefont {Q.}~\bibnamefont {Nguyen}}, \bibinfo {author} {\bibfnamefont
  {D.}~\bibnamefont {Kutnyakhov}}, \bibinfo {author} {\bibfnamefont
  {N.}~\bibnamefont {Wind}}, \bibinfo {author} {\bibfnamefont {L.}~\bibnamefont
  {Wenthaus}}, \bibinfo {author} {\bibfnamefont {M.}~\bibnamefont {Scholz}},
  \bibinfo {author} {\bibfnamefont {K.}~\bibnamefont {Rossnagel}}, \bibinfo
  {author} {\bibfnamefont {M.}~\bibnamefont {Hoesch}}, \bibinfo {author}
  {\bibfnamefont {M.}~\bibnamefont {Aeschlimann}}, \bibinfo {author}
  {\bibfnamefont {B.}~\bibnamefont {Stadtm{\"u}ller}}, \bibinfo {author}
  {\bibfnamefont {M.}~\bibnamefont {Kl{\"a}ui}}, \bibinfo {author}
  {\bibfnamefont {G.}~\bibnamefont {Sch{\"o}nhense}}, \bibinfo {author}
  {\bibfnamefont {T.}~\bibnamefont {Jungwirth}}, \bibinfo {author}
  {\bibfnamefont {A.~B.}\ \bibnamefont {Hellenes}}, \bibinfo {author}
  {\bibfnamefont {G.}~\bibnamefont {Jakob}}, \bibinfo {author} {\bibfnamefont
  {L.}~\bibnamefont {{\v S}mejkal}}, \bibinfo {author} {\bibfnamefont
  {J.}~\bibnamefont {Sinova}},\ and\ \bibinfo {author} {\bibfnamefont {H.-J.}\
  \bibnamefont {Elmers}},\ }\bibfield  {title} {\bibinfo {title} {Observation
  of time-reversal symmetry breaking in the band structure of altermagnetic
  {{RuO2}}},\ }\href {https://doi.org/10.1126/sciadv.adj4883} {\bibfield
  {journal} {\bibinfo  {journal} {Sci. Adv.}\ }\textbf {\bibinfo {volume}
  {10}},\ \bibinfo {pages} {eadj4883} (\bibinfo {year} {2024})}\BibitemShut
  {NoStop}%
\bibitem [{\citenamefont {Osumi}\ \emph {et~al.}(2024)\citenamefont {Osumi},
  \citenamefont {Souma}, \citenamefont {Aoyama}, \citenamefont {Yamauchi},
  \citenamefont {Honma}, \citenamefont {Nakayama}, \citenamefont {Takahashi},
  \citenamefont {Ohgushi},\ and\ \citenamefont {Sato}}]{Osumi_Phys.Rev.B_2024}%
  \BibitemOpen
  \bibfield  {author} {\bibinfo {author} {\bibfnamefont {T.}~\bibnamefont
  {Osumi}}, \bibinfo {author} {\bibfnamefont {S.}~\bibnamefont {Souma}},
  \bibinfo {author} {\bibfnamefont {T.}~\bibnamefont {Aoyama}}, \bibinfo
  {author} {\bibfnamefont {K.}~\bibnamefont {Yamauchi}}, \bibinfo {author}
  {\bibfnamefont {A.}~\bibnamefont {Honma}}, \bibinfo {author} {\bibfnamefont
  {K.}~\bibnamefont {Nakayama}}, \bibinfo {author} {\bibfnamefont
  {T.}~\bibnamefont {Takahashi}}, \bibinfo {author} {\bibfnamefont
  {K.}~\bibnamefont {Ohgushi}},\ and\ \bibinfo {author} {\bibfnamefont
  {T.}~\bibnamefont {Sato}},\ }\bibfield  {title} {\bibinfo {title}
  {Observation of a giant band splitting in altermagnetic {{MnTe}}},\ }\href
  {https://doi.org/10.1103/PhysRevB.109.115102} {\bibfield  {journal} {\bibinfo
   {journal} {Phys. Rev. B}\ }\textbf {\bibinfo {volume} {109}},\ \bibinfo
  {pages} {115102} (\bibinfo {year} {2024})}\BibitemShut {NoStop}%
\bibitem [{\citenamefont {Reimers}\ \emph {et~al.}(2024)\citenamefont
  {Reimers}, \citenamefont {Odenbreit}, \citenamefont {{\v S}mejkal},
  \citenamefont {Strocov}, \citenamefont {Constantinou}, \citenamefont
  {Hellenes}, \citenamefont {Jaeschke~Ubiergo}, \citenamefont {Campos},
  \citenamefont {Bharadwaj}, \citenamefont {Chakraborty}, \citenamefont
  {Denneulin}, \citenamefont {Shi}, \citenamefont {{Dunin-Borkowski}},
  \citenamefont {Das}, \citenamefont {Kl{\"a}ui}, \citenamefont {Sinova},\ and\
  \citenamefont {Jourdan}}]{Reimers_NatCommun_2024}%
  \BibitemOpen
  \bibfield  {author} {\bibinfo {author} {\bibfnamefont {S.}~\bibnamefont
  {Reimers}}, \bibinfo {author} {\bibfnamefont {L.}~\bibnamefont {Odenbreit}},
  \bibinfo {author} {\bibfnamefont {L.}~\bibnamefont {{\v S}mejkal}}, \bibinfo
  {author} {\bibfnamefont {V.~N.}\ \bibnamefont {Strocov}}, \bibinfo {author}
  {\bibfnamefont {P.}~\bibnamefont {Constantinou}}, \bibinfo {author}
  {\bibfnamefont {A.~B.}\ \bibnamefont {Hellenes}}, \bibinfo {author}
  {\bibfnamefont {R.}~\bibnamefont {Jaeschke~Ubiergo}}, \bibinfo {author}
  {\bibfnamefont {W.~H.}\ \bibnamefont {Campos}}, \bibinfo {author}
  {\bibfnamefont {V.~K.}\ \bibnamefont {Bharadwaj}}, \bibinfo {author}
  {\bibfnamefont {A.}~\bibnamefont {Chakraborty}}, \bibinfo {author}
  {\bibfnamefont {T.}~\bibnamefont {Denneulin}}, \bibinfo {author}
  {\bibfnamefont {W.}~\bibnamefont {Shi}}, \bibinfo {author} {\bibfnamefont
  {R.~E.}\ \bibnamefont {{Dunin-Borkowski}}}, \bibinfo {author} {\bibfnamefont
  {S.}~\bibnamefont {Das}}, \bibinfo {author} {\bibfnamefont {M.}~\bibnamefont
  {Kl{\"a}ui}}, \bibinfo {author} {\bibfnamefont {J.}~\bibnamefont {Sinova}},\
  and\ \bibinfo {author} {\bibfnamefont {M.}~\bibnamefont {Jourdan}},\
  }\bibfield  {title} {\bibinfo {title} {Direct observation of altermagnetic
  band splitting in {{CrSb}} thin films},\ }\href
  {https://doi.org/10.1038/s41467-024-46476-5} {\bibfield  {journal} {\bibinfo
  {journal} {Nature Commun.}\ }\textbf {\bibinfo {volume} {15}},\ \bibinfo
  {pages} {2116} (\bibinfo {year} {2024})}\BibitemShut {NoStop}%
\bibitem [{\citenamefont {Jiang}\ \emph {et~al.}(2025)\citenamefont {Jiang},
  \citenamefont {Hu}, \citenamefont {Bai}, \citenamefont {Song}, \citenamefont
  {Mu}, \citenamefont {Qu}, \citenamefont {Li}, \citenamefont {Zhu},
  \citenamefont {Pi}, \citenamefont {Wei}, \citenamefont {Sun}, \citenamefont
  {Huang}, \citenamefont {Zheng}, \citenamefont {Peng}, \citenamefont {He},
  \citenamefont {Li}, \citenamefont {Luo}, \citenamefont {Li}, \citenamefont
  {Chen}, \citenamefont {Li}, \citenamefont {Weng},\ and\ \citenamefont
  {Qian}}]{Jiang_Nat.Phys._2025}%
  \BibitemOpen
  \bibfield  {author} {\bibinfo {author} {\bibfnamefont {B.}~\bibnamefont
  {Jiang}}, \bibinfo {author} {\bibfnamefont {M.}~\bibnamefont {Hu}}, \bibinfo
  {author} {\bibfnamefont {J.}~\bibnamefont {Bai}}, \bibinfo {author}
  {\bibfnamefont {Z.}~\bibnamefont {Song}}, \bibinfo {author} {\bibfnamefont
  {C.}~\bibnamefont {Mu}}, \bibinfo {author} {\bibfnamefont {G.}~\bibnamefont
  {Qu}}, \bibinfo {author} {\bibfnamefont {W.}~\bibnamefont {Li}}, \bibinfo
  {author} {\bibfnamefont {W.}~\bibnamefont {Zhu}}, \bibinfo {author}
  {\bibfnamefont {H.}~\bibnamefont {Pi}}, \bibinfo {author} {\bibfnamefont
  {Z.}~\bibnamefont {Wei}}, \bibinfo {author} {\bibfnamefont {Y.-J.}\
  \bibnamefont {Sun}}, \bibinfo {author} {\bibfnamefont {Y.}~\bibnamefont
  {Huang}}, \bibinfo {author} {\bibfnamefont {X.}~\bibnamefont {Zheng}},
  \bibinfo {author} {\bibfnamefont {Y.}~\bibnamefont {Peng}}, \bibinfo {author}
  {\bibfnamefont {L.}~\bibnamefont {He}}, \bibinfo {author} {\bibfnamefont
  {S.}~\bibnamefont {Li}}, \bibinfo {author} {\bibfnamefont {J.}~\bibnamefont
  {Luo}}, \bibinfo {author} {\bibfnamefont {Z.}~\bibnamefont {Li}}, \bibinfo
  {author} {\bibfnamefont {G.}~\bibnamefont {Chen}}, \bibinfo {author}
  {\bibfnamefont {H.}~\bibnamefont {Li}}, \bibinfo {author} {\bibfnamefont
  {H.}~\bibnamefont {Weng}},\ and\ \bibinfo {author} {\bibfnamefont
  {T.}~\bibnamefont {Qian}},\ }\bibfield  {title} {\bibinfo {title} {A metallic
  room-temperature d-wave altermagnet},\ }\href
  {https://doi.org/10.1038/s41567-025-02822-y} {\bibfield  {journal} {\bibinfo
  {journal} {Nature Physics}\ }\textbf {\bibinfo {volume} {21}},\ \bibinfo
  {pages} {754} (\bibinfo {year} {2025})}\BibitemShut {NoStop}%
\bibitem [{\citenamefont {Papaj}(2023)}]{Papaj_Phys.Rev.B_2023}%
  \BibitemOpen
  \bibfield  {author} {\bibinfo {author} {\bibfnamefont {M.}~\bibnamefont
  {Papaj}},\ }\bibfield  {title} {\bibinfo {title} {Andreev reflection at the
  altermagnet-superconductor interface},\ }\href
  {https://doi.org/10.1103/PhysRevB.108.L060508} {\bibfield  {journal}
  {\bibinfo  {journal} {Phys. Rev. B}\ }\textbf {\bibinfo {volume} {108}},\
  \bibinfo {pages} {L060508} (\bibinfo {year} {2023})}\BibitemShut {NoStop}%
\bibitem [{\citenamefont {Amundsen}\ \emph {et~al.}(2024)\citenamefont
  {Amundsen}, \citenamefont {Brataas},\ and\ \citenamefont
  {Linder}}]{Amundsen_Phys.Rev.B_2024}%
  \BibitemOpen
  \bibfield  {author} {\bibinfo {author} {\bibfnamefont {M.}~\bibnamefont
  {Amundsen}}, \bibinfo {author} {\bibfnamefont {A.}~\bibnamefont {Brataas}},\
  and\ \bibinfo {author} {\bibfnamefont {J.}~\bibnamefont {Linder}},\
  }\bibfield  {title} {\bibinfo {title} {{{RKKY}} interaction in {{Rashba}}
  altermagnets},\ }\href {https://doi.org/10.1103/PhysRevB.110.054427}
  {\bibfield  {journal} {\bibinfo  {journal} {Phys. Rev. B}\ }\textbf {\bibinfo
  {volume} {110}},\ \bibinfo {pages} {054427} (\bibinfo {year}
  {2024})}\BibitemShut {NoStop}%
\bibitem [{\citenamefont {Banerjee}\ and\ \citenamefont
  {Scheurer}(2024)}]{Banerjee_Phys.Rev.B_2024}%
  \BibitemOpen
  \bibfield  {author} {\bibinfo {author} {\bibfnamefont {S.}~\bibnamefont
  {Banerjee}}\ and\ \bibinfo {author} {\bibfnamefont {M.~S.}\ \bibnamefont
  {Scheurer}},\ }\bibfield  {title} {\bibinfo {title} {Altermagnetic
  superconducting diode effect},\ }\href
  {https://doi.org/10.1103/PhysRevB.110.024503} {\bibfield  {journal} {\bibinfo
   {journal} {Phys. Rev. B}\ }\textbf {\bibinfo {volume} {110}},\ \bibinfo
  {pages} {024503} (\bibinfo {year} {2024})}\BibitemShut {NoStop}%
\bibitem [{\citenamefont {Lu}\ \emph {et~al.}(2024)\citenamefont {Lu},
  \citenamefont {Maeda}, \citenamefont {Ito}, \citenamefont {Yada},\ and\
  \citenamefont {Tanaka}}]{Lu_Phys.Rev.Lett._2024}%
  \BibitemOpen
  \bibfield  {author} {\bibinfo {author} {\bibfnamefont {B.}~\bibnamefont
  {Lu}}, \bibinfo {author} {\bibfnamefont {K.}~\bibnamefont {Maeda}}, \bibinfo
  {author} {\bibfnamefont {H.}~\bibnamefont {Ito}}, \bibinfo {author}
  {\bibfnamefont {K.}~\bibnamefont {Yada}},\ and\ \bibinfo {author}
  {\bibfnamefont {Y.}~\bibnamefont {Tanaka}},\ }\bibfield  {title} {\bibinfo
  {title} {$\varphi$ {{Josephson Junction Induced}} by {{Altermagnetism}}},\
  }\href {https://doi.org/10.1103/PhysRevLett.133.226002} {\bibfield  {journal}
  {\bibinfo  {journal} {Phys. Rev. Letters}\ }\textbf {\bibinfo {volume}
  {133}},\ \bibinfo {pages} {226002} (\bibinfo {year} {2024})}\BibitemShut
  {NoStop}%
\bibitem [{\citenamefont {Cheng}\ \emph {et~al.}(2024)\citenamefont {Cheng},
  \citenamefont {Mao},\ and\ \citenamefont {Sun}}]{Cheng_Phys.Rev.B_2024}%
  \BibitemOpen
  \bibfield  {author} {\bibinfo {author} {\bibfnamefont {Q.}~\bibnamefont
  {Cheng}}, \bibinfo {author} {\bibfnamefont {Y.}~\bibnamefont {Mao}},\ and\
  \bibinfo {author} {\bibfnamefont {Q.-F.}\ \bibnamefont {Sun}},\ }\bibfield
  {title} {\bibinfo {title} {Field-free {{Josephson}} diode effect in
  altermagnet/normal metal/altermagnet junctions},\ }\href
  {https://doi.org/10.1103/PhysRevB.110.014518} {\bibfield  {journal} {\bibinfo
   {journal} {Phys. Rev. B}\ }\textbf {\bibinfo {volume} {110}},\ \bibinfo
  {pages} {014518} (\bibinfo {year} {2024})}\BibitemShut {NoStop}%
\bibitem [{\citenamefont {Zhao}\ \emph {et~al.}(2025)\citenamefont {Zhao},
  \citenamefont {Fukaya}, \citenamefont {Burset}, \citenamefont {Cayao},
  \citenamefont {Tanaka},\ and\ \citenamefont {Lu}}]{Zhao_Phys.Rev.B_2025}%
  \BibitemOpen
  \bibfield  {author} {\bibinfo {author} {\bibfnamefont {W.}~\bibnamefont
  {Zhao}}, \bibinfo {author} {\bibfnamefont {Y.}~\bibnamefont {Fukaya}},
  \bibinfo {author} {\bibfnamefont {P.}~\bibnamefont {Burset}}, \bibinfo
  {author} {\bibfnamefont {J.}~\bibnamefont {Cayao}}, \bibinfo {author}
  {\bibfnamefont {Y.}~\bibnamefont {Tanaka}},\ and\ \bibinfo {author}
  {\bibfnamefont {B.}~\bibnamefont {Lu}},\ }\bibfield  {title} {\bibinfo
  {title} {Orientation-dependent transport in junctions formed by $d$-wave
  altermagnets and $d$-wave superconductors},\ }\href
  {https://doi.org/10.1103/PhysRevB.111.184515} {\bibfield  {journal} {\bibinfo
   {journal} {Phys. Rev. B}\ }\textbf {\bibinfo {volume} {111}},\ \bibinfo
  {pages} {184515} (\bibinfo {year} {2025})}\BibitemShut {NoStop}%
\bibitem [{\citenamefont {Chen}\ \emph {et~al.}(2025)\citenamefont {Chen},
  \citenamefont {Wu}, \citenamefont {Peng}, \citenamefont {Shao}, \citenamefont
  {Ouyang},\ and\ \citenamefont {Li}}]{Chen_Phys.Rev.B_2025}%
  \BibitemOpen
  \bibfield  {author} {\bibinfo {author} {\bibfnamefont {X.-Y.}\ \bibnamefont
  {Chen}}, \bibinfo {author} {\bibfnamefont {B.}~\bibnamefont {Wu}}, \bibinfo
  {author} {\bibfnamefont {J.-W.}\ \bibnamefont {Peng}}, \bibinfo {author}
  {\bibfnamefont {J.}~\bibnamefont {Shao}}, \bibinfo {author} {\bibfnamefont
  {G.}~\bibnamefont {Ouyang}},\ and\ \bibinfo {author} {\bibfnamefont
  {H.}~\bibnamefont {Li}},\ }\bibfield  {title} {\bibinfo {title}
  {Phase-coherent thermal transport in {{Josephson}} junctions based on
  altermagnets},\ }\href {https://doi.org/10.1103/PhysRevB.111.125405}
  {\bibfield  {journal} {\bibinfo  {journal} {Phys. Rev. B}\ }\textbf {\bibinfo
  {volume} {111}},\ \bibinfo {pages} {125405} (\bibinfo {year}
  {2025})}\BibitemShut {NoStop}%
\bibitem [{\citenamefont {Sim}\ and\ \citenamefont
  {Knolle}(2025)}]{Sim_Phys.Rev.B_2025}%
  \BibitemOpen
  \bibfield  {author} {\bibinfo {author} {\bibfnamefont {G.}~\bibnamefont
  {Sim}}\ and\ \bibinfo {author} {\bibfnamefont {J.}~\bibnamefont {Knolle}},\
  }\bibfield  {title} {\bibinfo {title} {Pair density waves and supercurrent
  diode effect in altermagnets},\ }\href {https://doi.org/10.1103/b7rh-v7nq}
  {\bibfield  {journal} {\bibinfo  {journal} {Phys. Rev. B}\ }\textbf {\bibinfo
  {volume} {112}},\ \bibinfo {pages} {L020502} (\bibinfo {year}
  {2025})}\BibitemShut {NoStop}%
\bibitem [{\citenamefont {Beenakker}\ and\ \citenamefont
  {Vakhtel}(2023)}]{Beenakker_Phys.Rev.B_2023}%
  \BibitemOpen
  \bibfield  {author} {\bibinfo {author} {\bibfnamefont {C.~W.~J.}\
  \bibnamefont {Beenakker}}\ and\ \bibinfo {author} {\bibfnamefont
  {T.}~\bibnamefont {Vakhtel}},\ }\bibfield  {title} {\bibinfo {title}
  {Phase-shifted {{Andreev}} levels in an altermagnet {{Josephson}} junction},\
  }\href {https://doi.org/10.1103/PhysRevB.108.075425} {\bibfield  {journal}
  {\bibinfo  {journal} {Phys. Rev. B}\ }\textbf {\bibinfo {volume} {108}},\
  \bibinfo {pages} {075425} (\bibinfo {year} {2023})}\BibitemShut {NoStop}%
\bibitem [{\citenamefont {Alipourzadeh}\ and\ \citenamefont
  {Hajati}(2025)}]{Alipourzadeh_Phys.Rev.B_2025}%
  \BibitemOpen
  \bibfield  {author} {\bibinfo {author} {\bibfnamefont {M.}~\bibnamefont
  {Alipourzadeh}}\ and\ \bibinfo {author} {\bibfnamefont {Y.}~\bibnamefont
  {Hajati}},\ }\bibfield  {title} {\bibinfo {title} {Andreev bound states and
  supercurrent in an unconventional superconductor-altermagnet {{Josephson}}
  junction},\ }\href {https://doi.org/10.1103/mj4b-2fnr} {\bibfield  {journal}
  {\bibinfo  {journal} {Phys. Rev. B}\ }\textbf {\bibinfo {volume} {111}},\
  \bibinfo {pages} {214515} (\bibinfo {year} {2025})}\BibitemShut {NoStop}%
\bibitem [{\citenamefont {Fukaya}\ \emph {et~al.}(2025)\citenamefont {Fukaya},
  \citenamefont {Maeda}, \citenamefont {Yada}, \citenamefont {Cayao},
  \citenamefont {Tanaka},\ and\ \citenamefont {Lu}}]{Fukaya_Phys.Rev.B_2025}%
  \BibitemOpen
  \bibfield  {author} {\bibinfo {author} {\bibfnamefont {Y.}~\bibnamefont
  {Fukaya}}, \bibinfo {author} {\bibfnamefont {K.}~\bibnamefont {Maeda}},
  \bibinfo {author} {\bibfnamefont {K.}~\bibnamefont {Yada}}, \bibinfo {author}
  {\bibfnamefont {J.}~\bibnamefont {Cayao}}, \bibinfo {author} {\bibfnamefont
  {Y.}~\bibnamefont {Tanaka}},\ and\ \bibinfo {author} {\bibfnamefont
  {B.}~\bibnamefont {Lu}},\ }\bibfield  {title} {\bibinfo {title} {Josephson
  effect and odd-frequency pairing in superconducting junctions with
  unconventional magnets},\ }\href
  {https://doi.org/10.1103/PhysRevB.111.064502} {\bibfield  {journal} {\bibinfo
   {journal} {Phys. Rev. B}\ }\textbf {\bibinfo {volume} {111}},\ \bibinfo
  {pages} {064502} (\bibinfo {year} {2025})}\BibitemShut {NoStop}%
\bibitem [{\citenamefont {Ouassou}\ \emph {et~al.}(2023)\citenamefont
  {Ouassou}, \citenamefont {Brataas},\ and\ \citenamefont
  {Linder}}]{Jabir_PRL_2023}%
  \BibitemOpen
  \bibfield  {author} {\bibinfo {author} {\bibfnamefont {J.~A.}\ \bibnamefont
  {Ouassou}}, \bibinfo {author} {\bibfnamefont {A.}~\bibnamefont {Brataas}},\
  and\ \bibinfo {author} {\bibfnamefont {J.}~\bibnamefont {Linder}},\
  }\bibfield  {title} {\bibinfo {title} {dc {{Josephson}} effect in
  {{Altermagnets}}},\ }\href {https://doi.org/10.1103/PhysRevLett.131.076003}
  {\bibfield  {journal} {\bibinfo  {journal} {Phys. Rev. Lett.}\ }\textbf
  {\bibinfo {volume} {131}},\ \bibinfo {pages} {076003} (\bibinfo {year}
  {2023})}\BibitemShut {NoStop}%
\bibitem [{\citenamefont {Giil}\ and\ \citenamefont
  {Linder}(2024)}]{Giil_Phys.Rev.B_2024}%
  \BibitemOpen
  \bibfield  {author} {\bibinfo {author} {\bibfnamefont {H.~G.}\ \bibnamefont
  {Giil}}\ and\ \bibinfo {author} {\bibfnamefont {J.}~\bibnamefont {Linder}},\
  }\bibfield  {title} {\bibinfo {title} {Superconductor-altermagnet memory
  functionality without stray fields},\ }\href
  {https://doi.org/10.1103/PhysRevB.109.134511} {\bibfield  {journal} {\bibinfo
   {journal} {Phys. Rev. B}\ }\textbf {\bibinfo {volume} {109}},\ \bibinfo
  {pages} {134511} (\bibinfo {year} {2024})}\BibitemShut {NoStop}%
\bibitem [{\citenamefont {Monkman}\ \emph {et~al.}(2026)\citenamefont
  {Monkman}, \citenamefont {Weng}, \citenamefont {Heinsdorf}, \citenamefont
  {Nocera}, \citenamefont {Barlas},\ and\ \citenamefont
  {Franz}}]{Monkman_Phys.Rev.X_2026}%
  \BibitemOpen
  \bibfield  {author} {\bibinfo {author} {\bibfnamefont {K.}~\bibnamefont
  {Monkman}}, \bibinfo {author} {\bibfnamefont {J.}~\bibnamefont {Weng}},
  \bibinfo {author} {\bibfnamefont {N.}~\bibnamefont {Heinsdorf}}, \bibinfo
  {author} {\bibfnamefont {A.}~\bibnamefont {Nocera}}, \bibinfo {author}
  {\bibfnamefont {Y.}~\bibnamefont {Barlas}},\ and\ \bibinfo {author}
  {\bibfnamefont {M.}~\bibnamefont {Franz}},\ }\bibfield  {title} {\bibinfo
  {title} {Persistent {{Spin Currents}} in {{Superconducting Altermagnets}}},\
  }\href {https://doi.org/10.1103/52wh-1z5y} {\bibfield  {journal} {\bibinfo
  {journal} {Phys. Rev. X}\ }\textbf {\bibinfo {volume} {16}},\ \bibinfo
  {pages} {011057} (\bibinfo {year} {2026})}\BibitemShut {NoStop}%
\bibitem [{\citenamefont {Zhu}\ \emph {et~al.}(2023)\citenamefont {Zhu},
  \citenamefont {Zhuang}, \citenamefont {Wu},\ and\ \citenamefont
  {Yan}}]{Zhu_Phys.Rev.B_2023}%
  \BibitemOpen
  \bibfield  {author} {\bibinfo {author} {\bibfnamefont {D.}~\bibnamefont
  {Zhu}}, \bibinfo {author} {\bibfnamefont {Z.-Y.}\ \bibnamefont {Zhuang}},
  \bibinfo {author} {\bibfnamefont {Z.}~\bibnamefont {Wu}},\ and\ \bibinfo
  {author} {\bibfnamefont {Z.}~\bibnamefont {Yan}},\ }\bibfield  {title}
  {\bibinfo {title} {Topological superconductivity in two-dimensional
  altermagnetic metals},\ }\href {https://doi.org/10.1103/PhysRevB.108.184505}
  {\bibfield  {journal} {\bibinfo  {journal} {Phys. Rev. B}\ }\textbf {\bibinfo
  {volume} {108}},\ \bibinfo {pages} {184505} (\bibinfo {year}
  {2023})}\BibitemShut {NoStop}%
\bibitem [{\citenamefont {Ghorashi}\ \emph {et~al.}(2024)\citenamefont
  {Ghorashi}, \citenamefont {Hughes},\ and\ \citenamefont
  {Cano}}]{Ghorashi_Phys.Rev.Lett._2024}%
  \BibitemOpen
  \bibfield  {author} {\bibinfo {author} {\bibfnamefont {S.~A.~A.}\
  \bibnamefont {Ghorashi}}, \bibinfo {author} {\bibfnamefont {T.~L.}\
  \bibnamefont {Hughes}},\ and\ \bibinfo {author} {\bibfnamefont
  {J.}~\bibnamefont {Cano}},\ }\bibfield  {title} {\bibinfo {title}
  {Altermagnetic {{Routes}} to {{Majorana Modes}} in {{Zero Net
  Magnetization}}},\ }\href {https://doi.org/10.1103/PhysRevLett.133.106601}
  {\bibfield  {journal} {\bibinfo  {journal} {Phys. Rev. Letters}\ }\textbf
  {\bibinfo {volume} {133}},\ \bibinfo {pages} {106601} (\bibinfo {year}
  {2024})}\BibitemShut {NoStop}%
\bibitem [{\citenamefont {Li}\ and\ \citenamefont
  {Liu}(2023)}]{Li_Phys.Rev.B_2023}%
  \BibitemOpen
  \bibfield  {author} {\bibinfo {author} {\bibfnamefont {Y.-X.}\ \bibnamefont
  {Li}}\ and\ \bibinfo {author} {\bibfnamefont {C.-C.}\ \bibnamefont {Liu}},\
  }\bibfield  {title} {\bibinfo {title} {Majorana corner modes and tunable
  patterns in an altermagnet heterostructure},\ }\href
  {https://doi.org/10.1103/PhysRevB.108.205410} {\bibfield  {journal} {\bibinfo
   {journal} {Phys. Rev. B}\ }\textbf {\bibinfo {volume} {108}},\ \bibinfo
  {pages} {205410} (\bibinfo {year} {2023})}\BibitemShut {NoStop}%
\bibitem [{\citenamefont {Sun}\ \emph {et~al.}(2025)\citenamefont {Sun},
  \citenamefont {Liu}, \citenamefont {Zhang},\ and\ \citenamefont
  {Qiao}}]{Sun_Phys.Rev.B_2025}%
  \BibitemOpen
  \bibfield  {author} {\bibinfo {author} {\bibfnamefont {H.-Y.}\ \bibnamefont
  {Sun}}, \bibinfo {author} {\bibfnamefont {L.}~\bibnamefont {Liu}}, \bibinfo
  {author} {\bibfnamefont {Y.-T.}\ \bibnamefont {Zhang}},\ and\ \bibinfo
  {author} {\bibfnamefont {Z.}~\bibnamefont {Qiao}},\ }\bibfield  {title}
  {\bibinfo {title} {Altermagnetism-induced iron-based third-order topological
  superconductivity},\ }\href {https://doi.org/10.1103/g17w-xs73} {\bibfield
  {journal} {\bibinfo  {journal} {Phys. Rev. B}\ }\textbf {\bibinfo {volume}
  {112}},\ \bibinfo {pages} {195402} (\bibinfo {year} {2025})}\BibitemShut
  {NoStop}%
\bibitem [{\citenamefont {Yi}\ \emph {et~al.}(2026)\citenamefont {Yi},
  \citenamefont {Mao}, \citenamefont {Miao},\ and\ \citenamefont
  {Sun}}]{Yi_Phys.Rev.B_2026}%
  \BibitemOpen
  \bibfield  {author} {\bibinfo {author} {\bibfnamefont {X.-J.}\ \bibnamefont
  {Yi}}, \bibinfo {author} {\bibfnamefont {Y.}~\bibnamefont {Mao}}, \bibinfo
  {author} {\bibfnamefont {C.-M.}\ \bibnamefont {Miao}},\ and\ \bibinfo
  {author} {\bibfnamefont {Q.-F.}\ \bibnamefont {Sun}},\ }\bibfield  {title}
  {\bibinfo {title} {Majorana modes in a helical altermagnet without net
  magnetism and spin-orbit coupling},\ }\href
  {https://doi.org/10.1103/gj6c-mcd2} {\bibfield  {journal} {\bibinfo
  {journal} {Phys. Rev. B}\ }\textbf {\bibinfo {volume} {113}},\ \bibinfo
  {pages} {L060408} (\bibinfo {year} {2026})}\BibitemShut {NoStop}%
\bibitem [{\citenamefont {Hadjipaschalis}\ \emph {et~al.}(2025)\citenamefont
  {Hadjipaschalis}, \citenamefont {Ghorashi},\ and\ \citenamefont
  {Cano}}]{Hadjipaschalis_Phys.Rev.B_2025}%
  \BibitemOpen
  \bibfield  {author} {\bibinfo {author} {\bibfnamefont {A.}~\bibnamefont
  {Hadjipaschalis}}, \bibinfo {author} {\bibfnamefont {S.~A.~A.}\ \bibnamefont
  {Ghorashi}},\ and\ \bibinfo {author} {\bibfnamefont {J.}~\bibnamefont
  {Cano}},\ }\bibfield  {title} {\bibinfo {title} {Majoranas with a twist:
  {{Tunable Majorana}} zero modes in altermagnetic heterostructures},\ }\href
  {https://doi.org/10.1103/p79l-rty6} {\bibfield  {journal} {\bibinfo
  {journal} {Phys. Rev. B}\ }\textbf {\bibinfo {volume} {112}},\ \bibinfo
  {pages} {214430} (\bibinfo {year} {2025})}\BibitemShut {NoStop}%
\bibitem [{\citenamefont {Mondal}\ \emph {et~al.}(2025)\citenamefont {Mondal},
  \citenamefont {Pal}, \citenamefont {Saha},\ and\ \citenamefont
  {Nag}}]{Mondal_Phys.Rev.B_2025}%
  \BibitemOpen
  \bibfield  {author} {\bibinfo {author} {\bibfnamefont {D.}~\bibnamefont
  {Mondal}}, \bibinfo {author} {\bibfnamefont {A.}~\bibnamefont {Pal}},
  \bibinfo {author} {\bibfnamefont {A.}~\bibnamefont {Saha}},\ and\ \bibinfo
  {author} {\bibfnamefont {T.}~\bibnamefont {Nag}},\ }\bibfield  {title}
  {\bibinfo {title} {Distinguishing between topological {{Majorana}} and
  trivial zero modes via transport and shot noise study in an altermagnet
  heterostructure},\ }\href {https://doi.org/10.1103/PhysRevB.111.L121401}
  {\bibfield  {journal} {\bibinfo  {journal} {Phys. Rev. B}\ }\textbf {\bibinfo
  {volume} {111}},\ \bibinfo {pages} {L121401} (\bibinfo {year}
  {2025})}\BibitemShut {NoStop}%
\bibitem [{\citenamefont {Pal}\ \emph {et~al.}(2025)\citenamefont {Pal},
  \citenamefont {Mondal}, \citenamefont {Nag},\ and\ \citenamefont
  {Saha}}]{Pal_Phys.Rev.B_2025}%
  \BibitemOpen
  \bibfield  {author} {\bibinfo {author} {\bibfnamefont {A.}~\bibnamefont
  {Pal}}, \bibinfo {author} {\bibfnamefont {D.}~\bibnamefont {Mondal}},
  \bibinfo {author} {\bibfnamefont {T.}~\bibnamefont {Nag}},\ and\ \bibinfo
  {author} {\bibfnamefont {A.}~\bibnamefont {Saha}},\ }\bibfield  {title}
  {\bibinfo {title} {Josephson current signature of {{Floquet Majorana}} and
  topological accidental zero modes in altermagnet heterostructures},\ }\href
  {https://doi.org/10.1103/prnx-47mk} {\bibfield  {journal} {\bibinfo
  {journal} {Phys. Rev. B}\ }\textbf {\bibinfo {volume} {112}},\ \bibinfo
  {pages} {L201408} (\bibinfo {year} {2025})}\BibitemShut {NoStop}%
\bibitem [{\citenamefont {Chatterjee}\ and\ \citenamefont {Juri{\v
  c}i{\'c}}(2025)}]{Chatterjee_Phys.Rev.B_2025}%
  \BibitemOpen
  \bibfield  {author} {\bibinfo {author} {\bibfnamefont {P.}~\bibnamefont
  {Chatterjee}}\ and\ \bibinfo {author} {\bibfnamefont {V.}~\bibnamefont
  {Juri{\v c}i{\'c}}},\ }\bibfield  {title} {\bibinfo {title} {Interplay
  between altermagnetism and topological superconductivity on an unconventional
  superconducting platform},\ }\href {https://doi.org/10.1103/4318-ttvf}
  {\bibfield  {journal} {\bibinfo  {journal} {Phys. Rev. B}\ }\textbf {\bibinfo
  {volume} {112}},\ \bibinfo {pages} {054503} (\bibinfo {year}
  {2025})}\BibitemShut {NoStop}%
\bibitem [{\citenamefont {Subhadarshini}\ \emph {et~al.}(2025)\citenamefont
  {Subhadarshini}, \citenamefont {Mishra},\ and\ \citenamefont
  {Saha}}]{Subhadarshini_Phys.Rev.B_2025}%
  \BibitemOpen
  \bibfield  {author} {\bibinfo {author} {\bibfnamefont {M.}~\bibnamefont
  {Subhadarshini}}, \bibinfo {author} {\bibfnamefont {A.}~\bibnamefont
  {Mishra}},\ and\ \bibinfo {author} {\bibfnamefont {A.}~\bibnamefont {Saha}},\
  }\bibfield  {title} {\bibinfo {title} {Engineering a second-order topological
  superconductor hosting tunable {{Majorana}} corner modes in a magnet/$d$-wave
  superconductor hybrid platform},\ }\href {https://doi.org/10.1103/npn5-g9kp}
  {\bibfield  {journal} {\bibinfo  {journal} {Phys. Rev. B}\ }\textbf {\bibinfo
  {volume} {112}},\ \bibinfo {pages} {125426} (\bibinfo {year}
  {2025})}\BibitemShut {NoStop}%
\bibitem [{\citenamefont {Yang}\ \emph {et~al.}(2025)\citenamefont {Yang},
  \citenamefont {Sun}, \citenamefont {Xie},\ and\ \citenamefont
  {Law}}]{Yang__2025}%
  \BibitemOpen
  \bibfield  {author} {\bibinfo {author} {\bibfnamefont {G.~Z.~X.}\
  \bibnamefont {Yang}}, \bibinfo {author} {\bibfnamefont {Z.-T.}\ \bibnamefont
  {Sun}}, \bibinfo {author} {\bibfnamefont {Y.-M.}\ \bibnamefont {Xie}},\ and\
  \bibinfo {author} {\bibfnamefont {K.~T.}\ \bibnamefont {Law}},\ }\href
  {https://doi.org/10.48550/arXiv.2502.20283} {\bibinfo {title} {Topological
  altermagnetic {{Josephson}} junctions}} (\bibinfo {year} {2025}),\ \Eprint
  {https://arxiv.org/abs/2502.20283} {arXiv:2502.20283 [cond-mat]} \BibitemShut
  {NoStop}%
\bibitem [{\citenamefont {Groth}\ \emph {et~al.}(2014)\citenamefont {Groth},
  \citenamefont {Wimmer}, \citenamefont {Akhmerov},\ and\ \citenamefont
  {Waintal}}]{Groth_NJP2014}%
  \BibitemOpen
  \bibfield  {author} {\bibinfo {author} {\bibfnamefont {C.~W.}\ \bibnamefont
  {Groth}}, \bibinfo {author} {\bibfnamefont {M.}~\bibnamefont {Wimmer}},
  \bibinfo {author} {\bibfnamefont {A.~R.}\ \bibnamefont {Akhmerov}},\ and\
  \bibinfo {author} {\bibfnamefont {X.}~\bibnamefont {Waintal}},\ }\bibfield
  {title} {\bibinfo {title} {Kwant: a software package for quantum transport},\
  }\href {https://doi.org/10.1088/1367-2630/16/6/063065} {\bibfield  {journal}
  {\bibinfo  {journal} {New J. Phys.}\ }\textbf {\bibinfo {volume} {16}},\
  \bibinfo {pages} {063065} (\bibinfo {year} {2014})}\BibitemShut {NoStop}%
\end{thebibliography}%

\end{document}